\documentclass[fleqn,usenatbib]{mnras}

\usepackage{txfonts}

\usepackage[T1]{fontenc}
\usepackage{ae,aecompl}

\usepackage{color}
\usepackage{subcaption}
\usepackage{booktabs}
\usepackage{captcont}
\usepackage[perpage]{footmisc}
\usepackage[colorinlistoftodos,prependcaption,textsize=footnotesize]{todonotes}
\usepackage{footmisc}
\usepackage{graphicx}	
\usepackage{amssymb}	
\usepackage{cleveref}

\DefineFNsymbols{mySymbols}{{\ensuremath\dagger}{\ensuremath\ddagger}\S\P
   *{**}{\ensuremath{\dagger\dagger}}{\ensuremath{\ddagger\ddagger}}}
\setfnsymbol{mySymbols}

\newcommand{\dualta}[1]{{\color{black} #1}}
\newcommand{\comment}[1]{{\color{black} #1}}
\newcommand{\jdot}{{$\dot{J}$}}
\newcommand{\mdot}{{$\dot{M}$}}

\title[The Solar Wind in Time II]{The Solar Wind in Time II: 3D stellar wind structure and radio emission}

\author[D. \'{O}~Fionnag\'{a}in et al.]
{D. \'{O} Fionnag\'{a}in$^{1}$\thanks{Email: ofionnad@tcd.ie}\href{ https://orcid.org/0000-0001-9747-3573}{\includegraphics[scale=0.1]{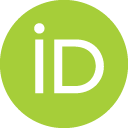}},
A.~A.~Vidotto$^{1}$,
P.~Petit$^{2,3}$,
C.~P. Folsom$^{2,3}$,
S.~V.~Jeffers$^{4}$ ,
\newauthor S.~C.~Marsden$^{5}$,
J.~Morin$^{6}$,
J.-D. do~Nascimento~Jr.$^{7,8}$,
\newauthor and the BCool Collaboration
\\
$^{1}$School of Physics, Trinity College Dublin, College Green, Dublin 2, Ireland\\
$^{2}$Universit\'{e} de Toulouse, UPS-OMP, Institut de Recherche en Astrophysique et Plan\'{e}tologie, Toulouse, France\\
$^{3}$IRAP, Universit\'e de Toulouse, CNRS, UPS, CNES, 31400, Toulouse, France\\
$^{4}$Universit\"{a}t G\"{o}ttingen, Institut f\"{u}r Astrophysik, Friedrich-Hund-Platz 1, 37077 G\"{o}ttingen, Germany\\
$^{5}$University of Southern Queensland, Centre for Astrophysics, Toowoomba, QLD, 4350, Australia\\
$^{6}$Laboratoire Univers et Particules de Montpellier, Universit\'{e} de Montpellier, CNRS, F-34095, France\\
$^{7}$Dep. de F\'{i}sica, Universidade Federal do Rio Grande do Norte, CEP: 59072-970 Natal, RN, Brazil\\
$^{8}$Harvard-Smithsonian Center for Astrophysics, Cambridge, MA 02138, USA\\
}

\date{Accepted XXX. Received YYY; in original form ZZZ}

\pubyear{2018}

\begin{document}
\label{firstpage}
\pagerange{\pageref{firstpage}--\pageref{lastpage}}
\maketitle

\begin{abstract}
In this work, we simulate the evolution of the solar wind along its main sequence lifetime and compute its thermal radio emission. To study the evolution of the solar wind, we use a sample of solar mass stars at different ages. All these stars have observationally-reconstructed magnetic maps, which are incorporated in our 3D magnetohydrodynamic simulations of their winds. We show that angular-momentum loss and mass-loss rates decrease steadily on evolutionary timescales, although they can vary in a magnetic cycle timescale. Stellar winds are known to emit radiation in the form of thermal bremsstrahlung in the radio spectrum. To calculate the expected radio fluxes from these winds, we solve the radiative transfer equation numerically from first principles. We compute continuum spectra across the frequency range 100 MHz - 100 GHz and find maximum radio flux densities ranging from 0.05 - 8.3 $\mu$Jy. At a frequency of 1 GHz and a normalised distance of d = 10 pc, the radio flux density follows 0.24 $(\Omega/\Omega_{\odot})^{0.9}$ (d/[10pc])$^2$ $\mu$Jy, where $\Omega$ is the rotation rate. This means that the best candidates for stellar wind observations in the radio regime are faster rotators within distances of 10 pc, such as $\kappa^1$ Ceti (2.83 $\mu$Jy) and $\chi^1$ Ori (8.3 $\mu$Jy). These flux predictions provide a guide to observing solar-type stars across the frequency range 0.1 - 100 GHz in the future using the next generation of radio telescopes, such as ngVLA and SKA. \href{https://papergit.page.link/3dwinds}{\includegraphics[width=10pt]{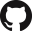}} \href{https://papergit.page.link/zenodo}{\includegraphics[width=11pt]{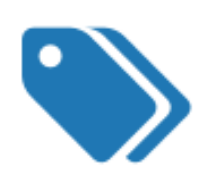}} 
\end{abstract}




\section{Introduction}
\label{sec:intro}
Solar analogues are essential to our understanding of how our own Sun has evolved through its past and how it will evolve into the future. The rotational evolution of stars has a significant effect on the activity \citep{Wright2011,Vidotto2014a}, as rotation has been linked to activity markers such as coronal X-ray emission \citep{Telleschi2005,Wright2011}, chromospheric activity (e.g. CaII, H$\alpha$) \citep{Lorenzo-Oliveira2018} and flaring rates \citep{Maehara2017}. \dualta{The stellar dynamo is regulated by rotation and convection, which in turn generates the magnetic field causing stellar activity \citep{Brun2017}.} By virtue of this relationship between rotation and activity, the  evolution of orbiting planets is directly affected, e.g. by high energy stellar radiation incident on their atmospheres \citep{Ribas2016,Owen2016}. Stellar rotation has been shown to decrease with age \citep{Skumanich1972} following $\Omega \propto t^{1/2}$ for stars older than $\sim 700$~Myr \citep{Gallet2013}. More recently, however, some deviation from this standardised age-rotation relationship has been observed at older ages \citep{VanSaders2016}, with some processes proposed to explain this behaviour \citep{Metcalfe2016,Beck2017,Booth2017,OFionnagain2018}. \par
The mechanism by which stars spin down while traversing the main sequence is through angular momentum loss  by their magnetised winds (e.g. \citealt{Weber1967,Vidotto2014b,See2017}). Therefore, this indicates that the surface magnetic field of the star also evolves with time,  \dualta{as demonstrated with magnetic field observations analysed using the Zeeman-Doppler Imaging (ZDI) technique \citep{Vidotto2014a,Folsom2016,Folsom2018d}. ZDI is a method that allows for the reconstruction of the large-scale magnetic field of the stellar surface from a set of high-resolution spectropolarimetric data \citep{Semel1989,Brown1991,Donati1997}, although it is insensitive to small-scale fields (\citealt{Lang2014}, Lehmann et al. subm.)}. \citet{See2017b, See2017}  determined, from 66 ZDI-observed stars, that the magnetic geometry as well as angular momentum and mass loss is correlated to Rossby number\footnote{Rossby number (Ro) is defined as the ratio between stellar rotation and convective turnover time. \citep{Noyes1984}}. Other works have demonstrated that there is a link between all of stellar activity, magnetic strength and geometry, angular momentum loss, and stellar winds \citep{Nicholson2016,Matt2012,Pantolmos2017a,Finley2018}.

\dualta{Stellar angular momentum-loss depends upon how much mass is lost by their winds \citep{Weber1967}. Due to the tenuous nature of low-mass stellar winds, a direct measurement of their winds is difficult (e.g. \citealt{Wood2005}), but would prove extremely useful in the constraining of mass-loss rates and other global wind parameters.} In this regard, the observations of radio emission from the winds of low-mass stars could provide meaningful constraints on wind density and mass-loss rate \citep{Lim1996,Gudel2002,Villadsen2014,Fichtinger2017,Vidotto2017a}. The wind is expected to have continuum emission in radio through the mechanism of thermal free-free emission \citep{Panagia1975,Wright1975}. This emission is expected to be stronger for stars with denser winds and is also dependent on the density (n) gradient in the wind with radial distance, $R$: n $\propto R^{-a}$. The value of $a$ is indirectly related to other stellar parameters such as the specific gravity, magnetic field and rotation. When $a = 2$ this represents when the wind has reached terminal radial velocity, however, this is unrealistic in regions closer to the star where the wind is accelerating. Therefore, we expect stellar winds to exhibit gradients much steeper than when $a = 2$. We discuss this further in \Cref{sec:radio}.\par
With this idea in mind, \citet{Gudel1998} and \citet{Gaidos2000} observed various solar analogues. They could place upper limits on the radio fluxes from these objects, and so indirectly infer upper mass-loss rate constraints.  All non-degenerate stars emit some form of radio emission from their atmospheres \citep{Gudel2002}. Although different radio emission mechanisms dominate at different layers in their atmosphere and wind \citep{Gudel2002}. For example, detecting coronal radio flares at a given frequency implies the surrounding wind is optically thin at those frequencies, allowing for placement of upper mass-loss limits. In addition, \citet{Gudel2007a} noted that thermal emission should dominate at radio frequencies as long as no flares occur while observing. The three dominant thermal emission mechanisms the author described are bremsstrahlung from the chromosphere, cyclotron emission above active regions, and coronal bremsstrahlung from hot coronal loops. These emission mechanisms must be addressed when attempting to detect the winds of solar-type stars at radio frequencies.
\renewcommand*{\thefootnote}{\alph{footnote}}
 \begin{table*}
 \centering
 \begin{minipage}{\textwidth}
 \centering
  \caption{Stellar parameters of our sample are shown on the left (mass, radius, rotation period, age, and distance) and specifics of the simulations are shown on the right (base density, base temperature, mass-loss rate, angular momentum-loss rate, open magnetic flux, and flux ratio between surface and open magnetic fluxes). Stellar parameters were compiled in \citet{Vidotto2014a}. Distances are found using the Gaia DR2 database\protect\footnotemark[1] \citep{GaiaCollaboration2016,Brown2018} values for parallax.}
  \label{tab:general}
 \begin{tabular}{lcccccccccccc}
 \hline
  & \multicolumn{6}{c}{Observables} & \multicolumn{6}{c}{Simulation} \\
  \cmidrule(lr){2-7}\cmidrule(lr){8-13}
 Star & M$_\star$ & R$_\star$ & P$_{\rm rot}$ & $\Omega$ & Age & d & $n_0$ (cm$^{-3}$) & $T_0$ & $\dot{M}$ (M$_{\odot}$/yr) & $\dot{J}$ (ergs) & $\Phi_{\rm open}$ (G cm) & f\\
  & (M$_\odot$) & (R$_\odot$) & (d) & ($\Omega_{\odot}$) & (Gyr) & (pc) & ($\times10^8$) & (MK)  & ($\times10^{-13}$) & ($\times10^{30}$) & ($\times10^{22}$) & \\
 \hline \hline
 $\chi^1$ Ori & 1.03 & 1.05 & 4.86 & 5.60 & 0.5 & 8.84$^{\pm0.02}$ & 18.9 & 2.84 & 46.5 & 285 & 22.5 & 0.37\\
 HD 190771 & 0.96 & 0.98 & 8.8 & 3.09 & 2.7 & 19.02$^{\pm0.01}$ & 13.2 & 3.04 & 36.1 & 91.0 & 23.46 & 0.59\\
 $\kappa^1$ Ceti & 1.03 & 0.95 & 9.3 & 2.92 & 0.65 & 9.15$^{\pm0.03}$ & 12.8 & 2.98 & 22.1 & 124 & 30.71 & 0.44\\
 HD 76151 & 1.06 & 0.98 & 15.2 & 1.79 & 3.6 & 16.85$^{\pm0.01}$ & 9.54 & 2.47 & 8.26 & 31.8 & 14.68 & 0.49\\
 18 Sco & 0.98 & 1.02 & 22.7 & 1.20 & 3.0 & 14.13$^{\pm0.02}$ & 7.5 & 1.85 & 6.47 & 5.34 & 4.29 & 0.70\\
 HD 9986 & 1.02 & 1.04 & 23 & 1.18 & 4.3 & 25.46$^{\pm0.03}$ & 7.44 & 1.82 & 5.82 & 2.35 & 3.30 & 0.94\\
 Sun Min & 1.0 & 1.0 & 27.2 & 1 & 4.6 & - & 6.72 & 1.5 & 1.08 & 1.04 & 3.44 & 0.69\\
 Sun Max & 1.0 & 1.0 & 27.2 & 1 & 4.6 & - & 6.72 & 1.5 & 1.94 & 15.5 & 6.17 & 0.24\\
 \hline
 \vspace{-1cm}
 \footnotetext[1]{https://gea.esac.esa.int/archive/}
 \end{tabular}
 \end{minipage}
 \end{table*}
 \renewcommand*{\thefootnote}{\arabic{footnote}}
\par 
Observing these winds can become difficult as the fluxes expected from these sources is at the $\mu$Jy level (see upper limits placed by \citealt{Gaidos2000,Villadsen2014,Fichtinger2017}), \dualta{and can be drowned out by chromospheric and coronal emission as described in the previous paragraph}. \citet{Villadsen2014} observed three low-mass stars, with positive detections for all three stars in the Ku band (centred at 34.5 GHz) of the VLA, and non-detections at lower frequencies. They suggested that the detected emissions originate in the chromosphere of these stars, with some contributions from other sources of radio emission. If emanating from the chromosphere, these detections do not aid in constraining the wind. \citet{Fichtinger2017} more recently observed four solar-type stars with the VLA at radio frequencies, and provided  upper limits to the mass-loss rates for each, ranging from $3\times10^{-12}\ -\ 7\times10^{-10} M_{\odot}$ yr$^{-1}$, depending on how collimated the winds are. \citet{Bower2016} observed radio emission from the young star V830 Tau, with which \citet{Vidotto2017a} were able to propose mass-loss rate constraints between $3\times10^{-10}$ and $3\times10^{-12}$ $M_{\odot}$ yr$^{-1}$. Transient CMEs should also be observable, which would cause more issues in detecting the ambient stellar wind, but these events are expected to be relatively short and could also help in constraining transient mass-loss from these stars \citep{Crosley2016}.

To aid in the radio detection and interpretation of the winds of solar-type stars, we here quantify the detectability of the winds of 6 solar-like stars of different ages within the radio regime from 100 MHz - 100 GHz. We aim to study the effects ageing stellar winds have on different solar analogues along the main sequence, allowing us to constrain global parameters and quantify the local wind environment.
\begin{figure*}
        \centering
    \includegraphics[width=\hsize,keepaspectratio]{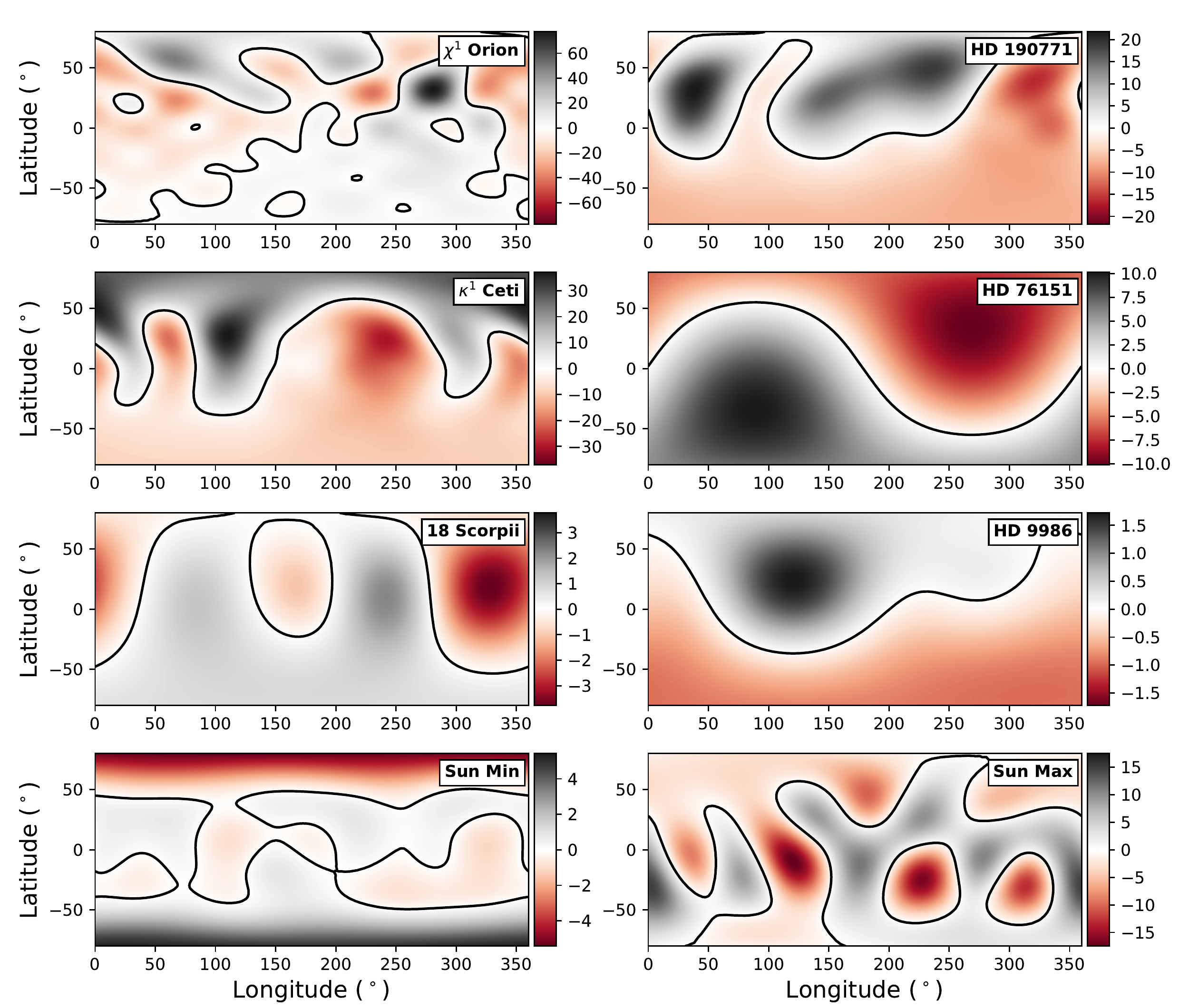}
    \caption{Radial surface magnetic fields of our stars. Each magnetic field is saturated at the maximum absolute value for each field respectively. Magnetic field contours are shown in Gauss. The maps are shown in latitude-longitude coordinates.
              }
         \label{fig:bfields}
\end{figure*}
To do this, we conduct 3D magnetohydrodynamical simulations of winds of solar-type stars, investigating the main-sequence solar wind evolution in terms of angular-momentum loss rates (\jdot), mass-loss rates (\mdot) and wind structure. We then use the results of our simulations to quantify the detectability of the radio emission from the solar wind in time, that can help guiding and planning of future observations of solar-like winds.
We present the sample of stars simulated and analysed in \Cref{sec:stellarsample}. Discussed in \Cref{sec:modelling} is the stellar wind modelling and simulation results. Our models predict the evolution of \jdot, \mdot, and $\Phi_{\rm open}$ of the solar wind through time, while also constraining the planetary environment surrounding the host stars. In \Cref{sec:radio} we demonstrate how we calculate radio emission for each star and the resulting emissions and flux densities expected. \Cref{sec:conclusion} we conclude on the results presented in this work. 
\section{Stellar Sample}
\label{sec:stellarsample}
Our sample of solar-like stars was selected so as to closely resemble to Sun in both mass and radius. \dualta{They cover a range of rotation rates (from 4.8-27 days or 1-5.6 $\Omega_{\odot}$) with ZDI reconstructed by \citet{Petit2008,DoNascimentoJr.2016} and Petit et al. (in prep) as part of the BCool collaboration}. \dualta{\citet{Gallet2013,Gallet2015a} depict different age-rotation evolutionary tracks for a 1 M$_{\odot}$ star, which converge at 800 Myr to the Skumanich law \citep{Skumanich1972}. $\chi^1$ Ori follows the fast rotator track, while the rest of our stars exist beyond the convergence point. We note that HD 190771 and HD 76151 exhibit faster rotation than the Skumanich law, which could be due to uncertainties in their ages.} The stars in our sample are listed below, see also \Cref{tab:general} for stellar parameters, and \Cref{fig:bfields} for observed ZDI maps.
 \begin{description}
 \item[\textbf{$\chi^1$ Orion}]{This star is both the youngest star and the fastest rotator we have simulated, with a rotation period of 4.8 days and an age of 0.5 Gyr \citep{Vidotto2014a}. This fast rotation should indicate a more active star than the slower rotators, which we see confirmed in the high magnetic field strengths. \dualta{The large-scale magnetic geometry reconstructed with ZDI for this star displays a complex structure (\Cref{fig:bfields}), showing very un-dipolar like structure (Petit et al., in prep.).} \dualta{Note that the ZDI observations here include 10 spherical harmonic degrees, which is the most of all simulations.} This star is the closest star in our sample at 8.84 pc\footnote{https://gea.esac.esa.int/archive/\label{note1}}.}

 \item[\textbf{HD 190771}]{This star possesses an uncharacteristically short rotation period (8.8 days) for its commonly used age (2.7 Gyr, derived from isochrone fitting, \citealt{Valenti2005}). This fast rotation should indicate a more active star, which we see validated in the ZDI observations of the magnetic field at the stellar surface. We see one of the least dipolar fields in the sample, with large areas of strong magnetic field of both polarities in the northern hemisphere (\Cref{fig:bfields}). \dualta{Note that polarity reversal has been observed to occur in the magnetic field of this star \citep{Petit2009}. } }

 \item[\textbf{$\kappa^1$ Ceti}]{is estimated to be the second youngest star in our selected sample, with an age of 0.65 Gyr \citep{Rosen2016}. \dualta{The observed rotation period from photometry is 9.2 days (\citealt{Messina2003,Rucinski2004}, ground and space respectively)}. The higher levels of activity in this star are apparent when we examine the ZDI map, with non-dipolar geometry and relatively strong B field (B$_{r,max} \approx$ 35 G,  \citealt{DoNascimentoJr.2016}). It is the second closest star in our sample (excluding the Sun), at a distance of 9.13 pc\textsuperscript{\ref{note1}}.}

 \item[\textbf{HD 76151}]{has a rotation period of 15.2 days \citep{Maldonado2010}. The age of HD 76151 is estimated to be 3.6 Gyr \citep{Petit2008}. ZDI observations of HD 76151 present a strong dipolar field, with B$_{\rm r,max}\approx$ 10 G, which is tilted to the axis of rotation by 30$^{\circ}$ \citep{Petit2008}. Considering the age of the star and the dipolar geometry of the magnetic field, we expect a slower wind than the faster, more magnetically active rotators.}

 \item[\textbf{18 Scorpii}]{ is 3 Gyr old and possesses a rotation period of 22.3 days. It displays very quiescent behaviour, with a weak, largely dipolar magnetic field \citep{Petit2008}. \dualta{It is the most similar solar twin for which we have surface magnetic field measurements, displaying very similar spectral lines to the Sun \citep{Melendez2014}. Recently, many more solar twins have been identified \citep{Lorenzo-Oliveira2018}, however, these stars do not have magnetic field observations. }}

 \item[\textbf{HD 9986}]{ presents another off axis dipole, with a maximum field strength of 1.6 G and an age of 4.3 Gyr \citep{Vidotto2014a}. This is the weakest magnetic field of any star in the sample, Petit et al. (in prep.)}

 \item[\textbf{The Sun}]{ has a well documented cyclical behaviour, of which we take one map at the maximum of the cycle, and another map at the minimum of the cycle. Maps for the minima and maxima are taken at Carrington rotations 1983 and 2078 respectively, which were observed with SOHO/MDI in the years 2001 and 2008. We have removed the higher degree harmonics ($\ell \geq 5$) for both maps, so as to replicate the Sun as if observed similarly to the other slowly rotating stars in the sample \citep{Vidotto2016,Vidotto2018,Lehmann2018}. We note that the Sun at maximum possesses a much more complex magnetic geometry than the solar minimum, including a stronger magnetic field (e.g. \citealt{DeRosa2010}).}
 \end{description}
  \begin{figure*}
    \centering
    \begin{subfigure}[b]{.43\linewidth}
    \includegraphics[width=\linewidth]{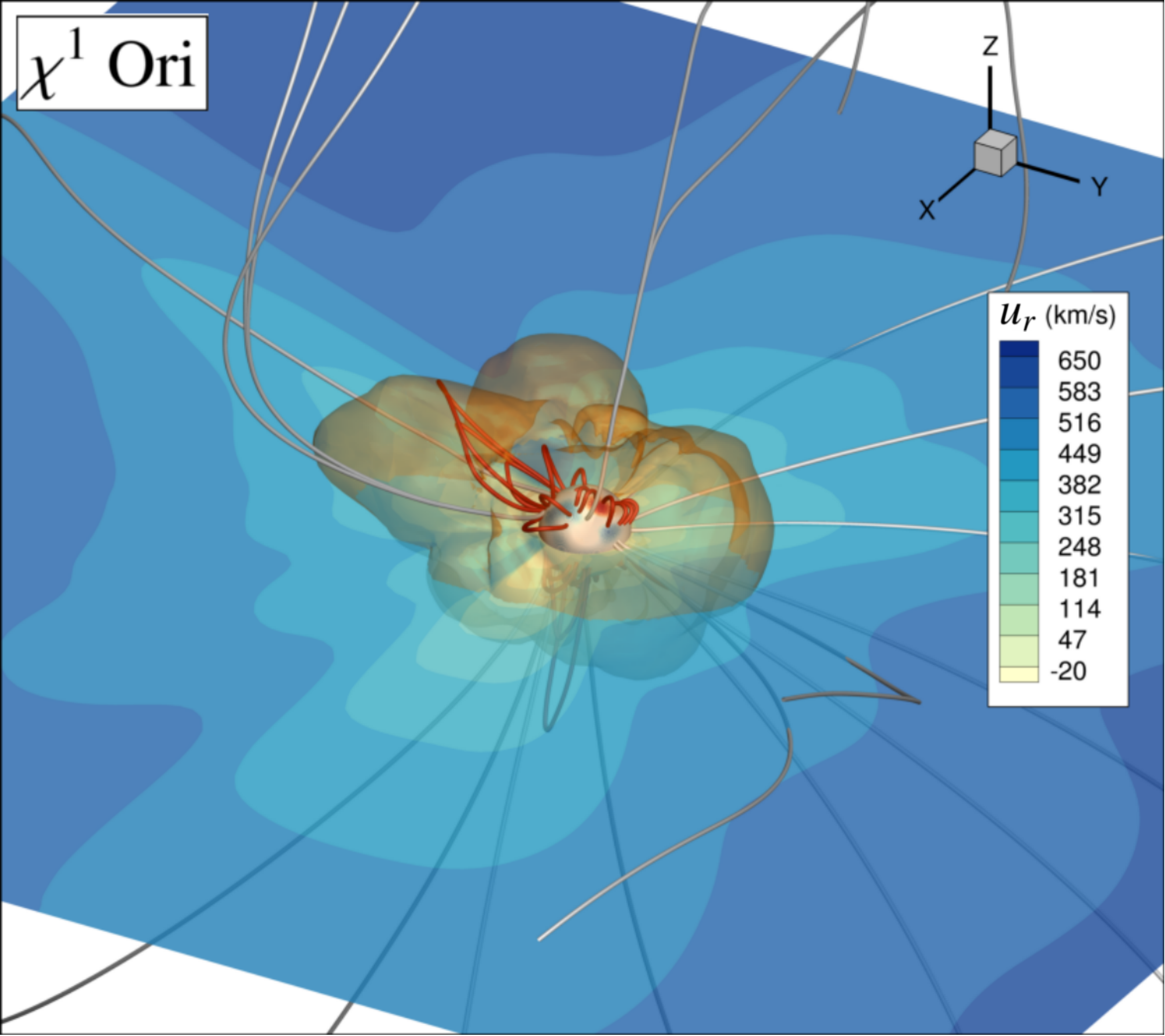}
    \label{fig:x1ori}
    \end{subfigure}
    \begin{subfigure}[b]{.43\linewidth}
    \includegraphics[width=\linewidth]{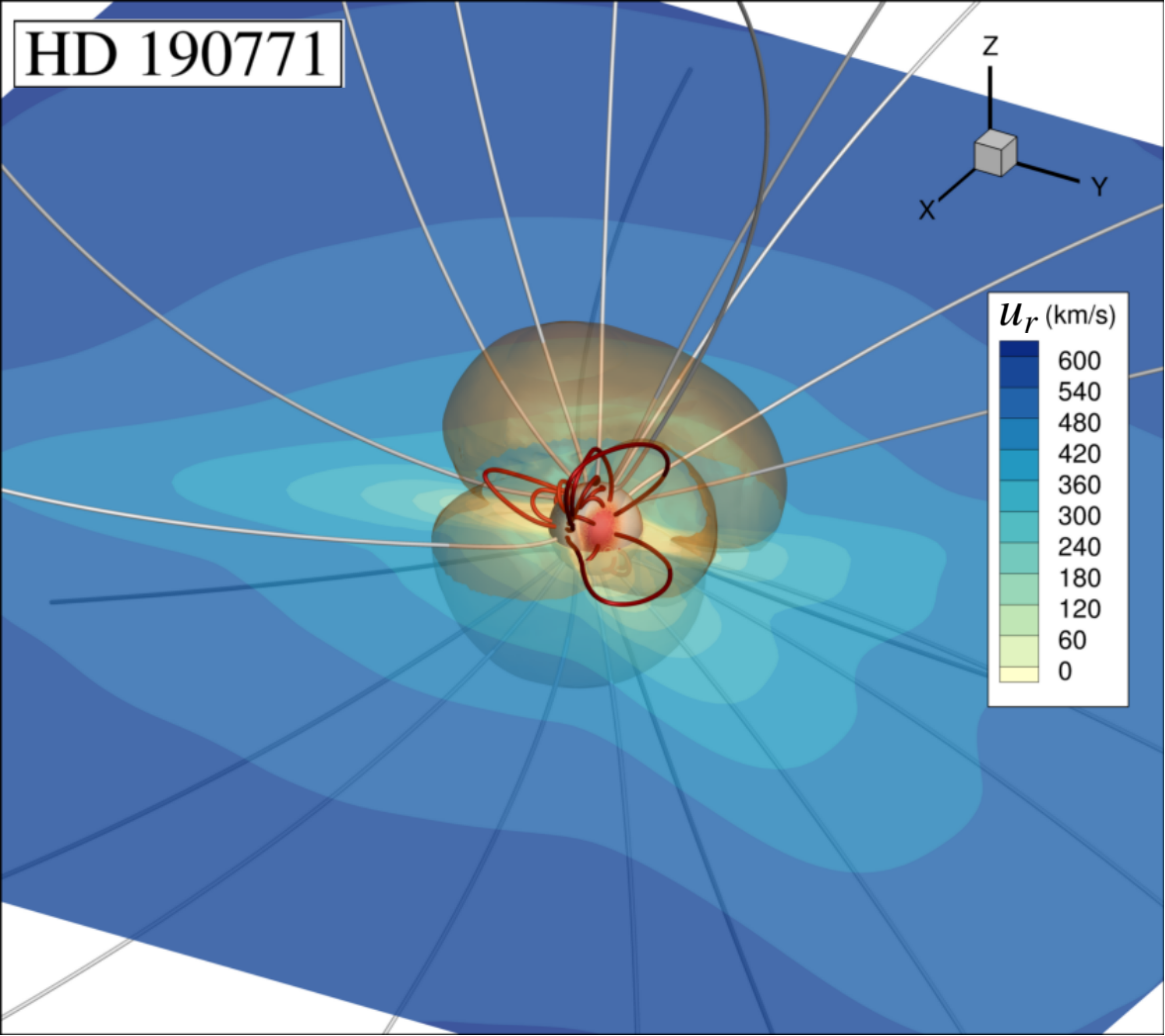}
    \label{fig:hd190771}
    \end{subfigure}

    \begin{subfigure}[b]{.43\linewidth}
    \includegraphics[width=\linewidth]{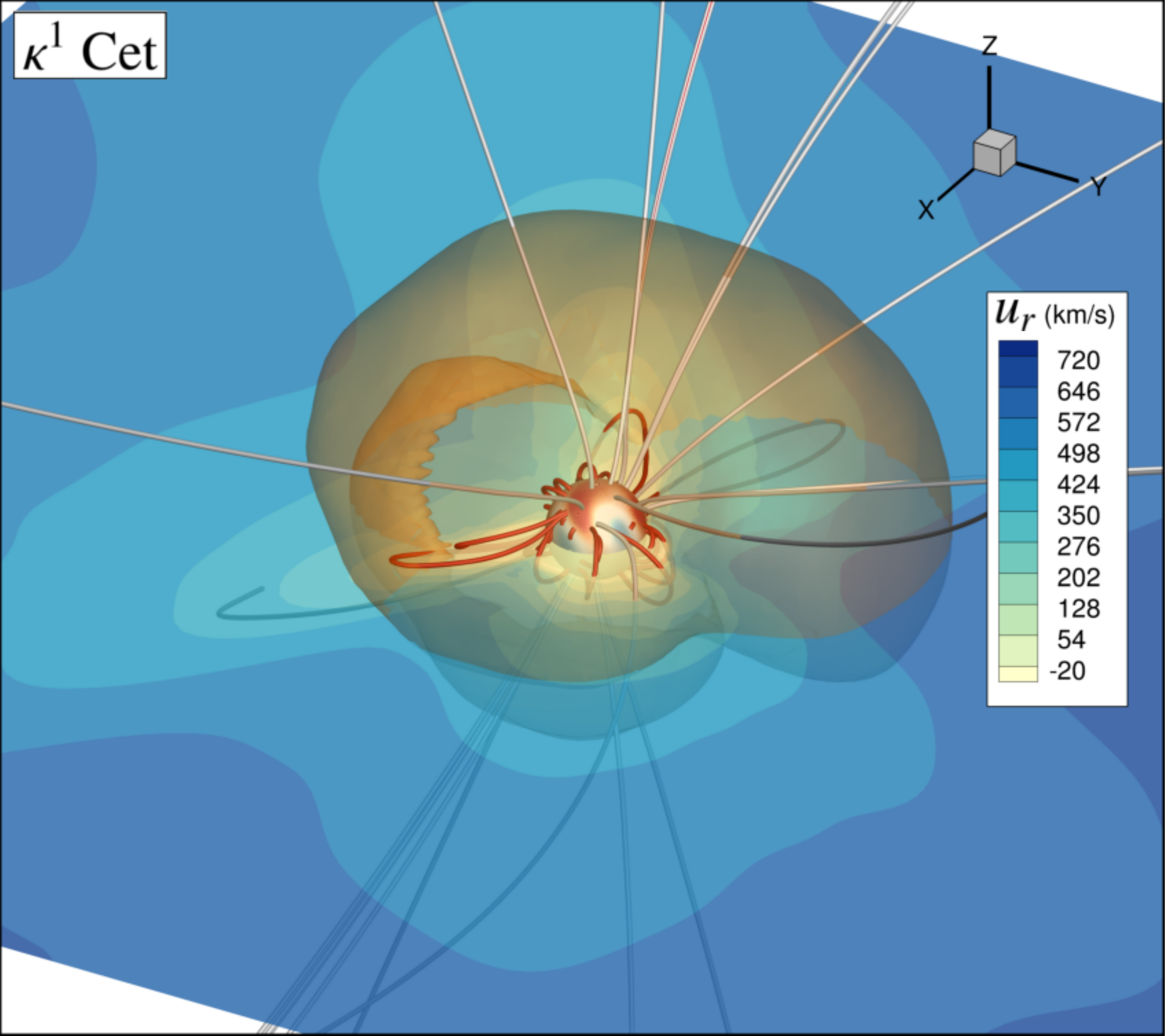}
    \label{fig:k1ceti}
    \end{subfigure}
    \begin{subfigure}[b]{.43\linewidth}
    \includegraphics[width=\linewidth]{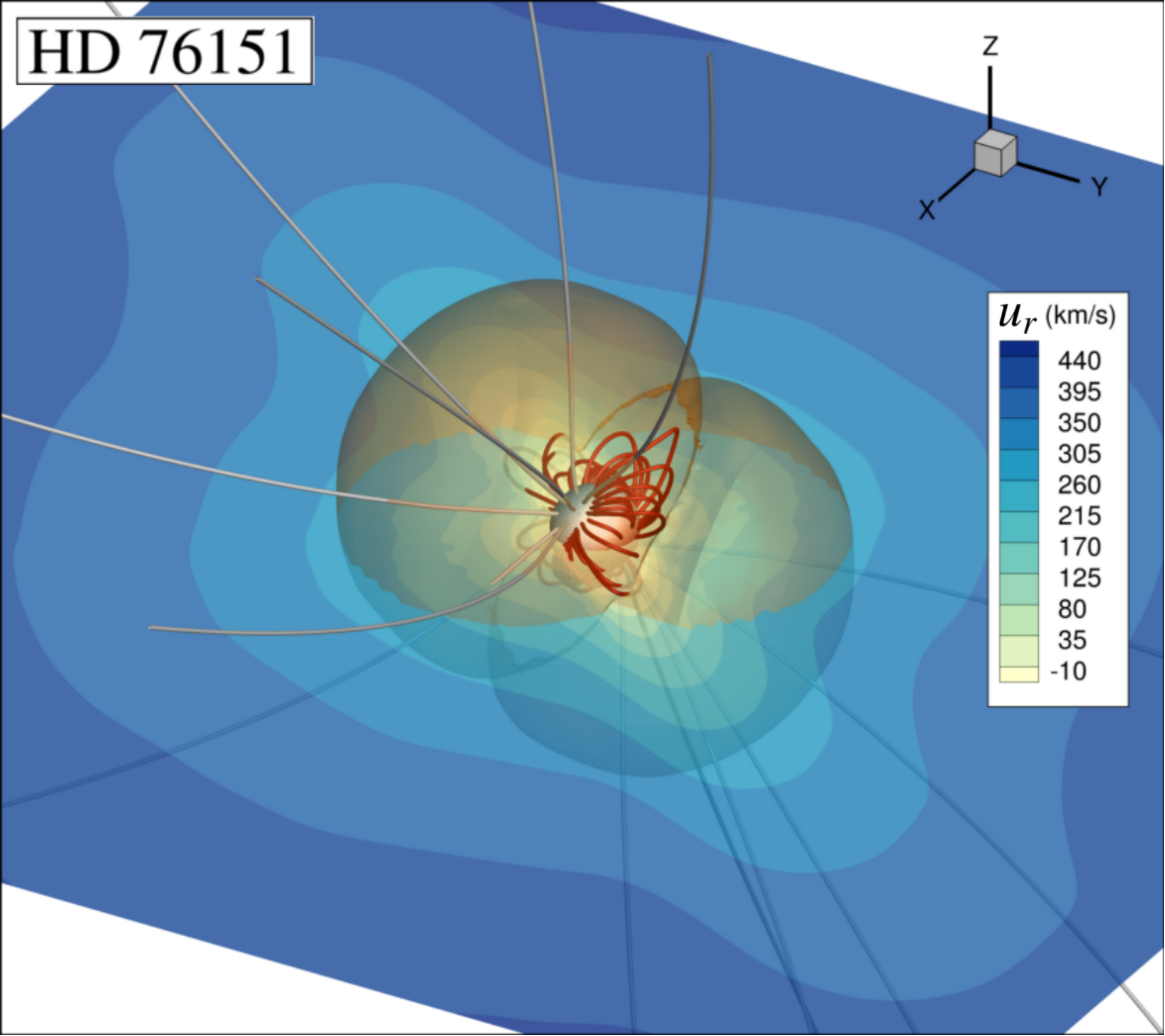}
    \label{fig:hd76151}
    \end{subfigure}
    \caption{Steady state solutions for the simulated winds of the solar analogues. The translucent slice through the z=0 plane shows the wind radial velocity ($u_r$). Open and closed magnetic field lines are shown as grey and red streamlines respectively. Magnetic polarity is shown on the stellar surface as a red-blue diverging contour. The orange surface shows the Alfv\'{e}n surface, where $u_r = u_A$, the Alfv\'{e}n velocity. Note that the faster rotators have much less uniform, dipolar Alfv\'{e}n surfaces, due to the less uniform magnetic fields topologically, at their surfaces.}
    \label{fig:tecplot1}
    \end{figure*}

    \begin{figure*} \ContinuedFloat
    \centering
    \begin{subfigure}[b]{.43\linewidth}
    \includegraphics[width=\linewidth]{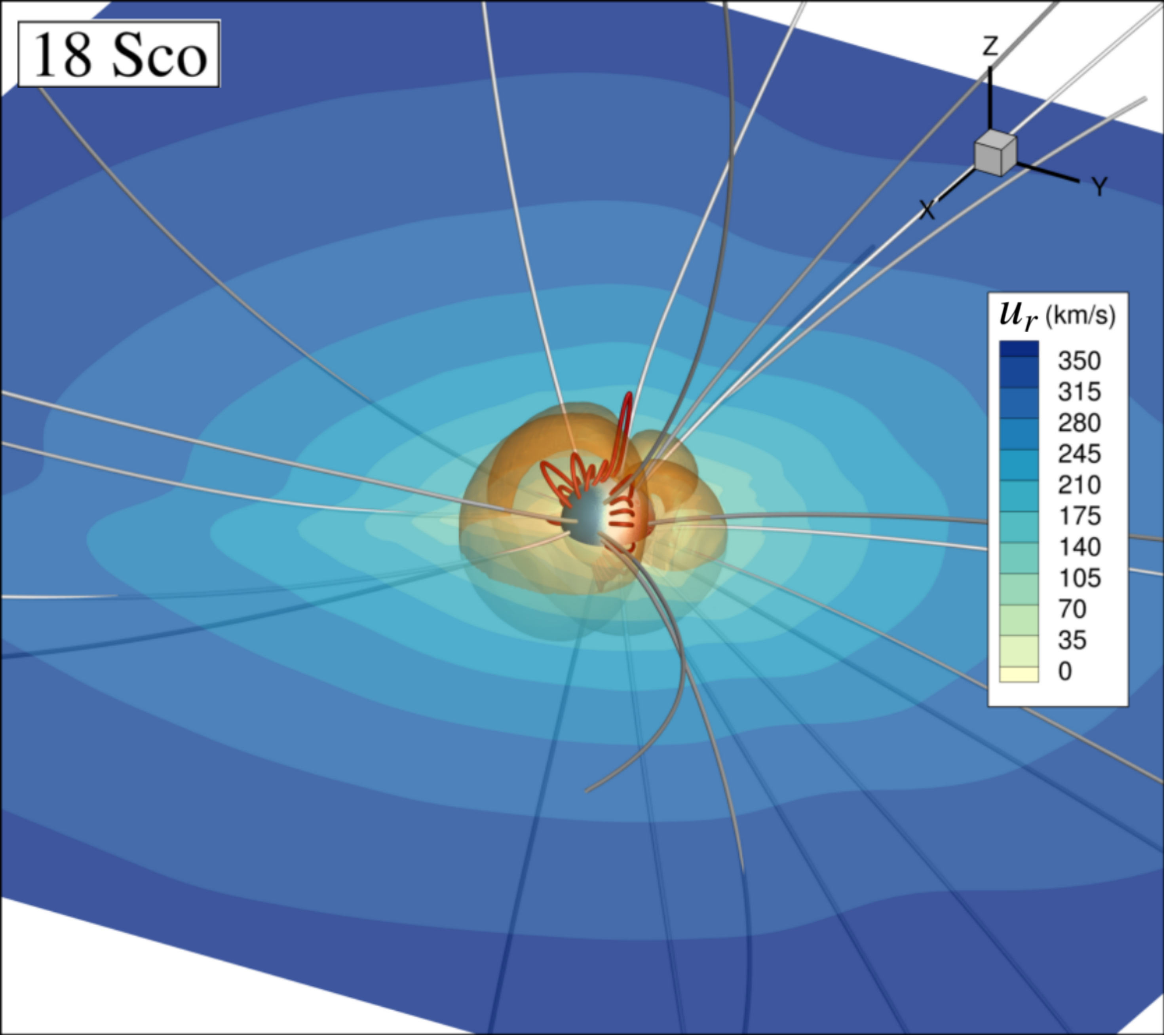}
    \label{fig:18sco}
    \end{subfigure}
    \begin{subfigure}[b]{.43\linewidth}
    \includegraphics[width=\linewidth]{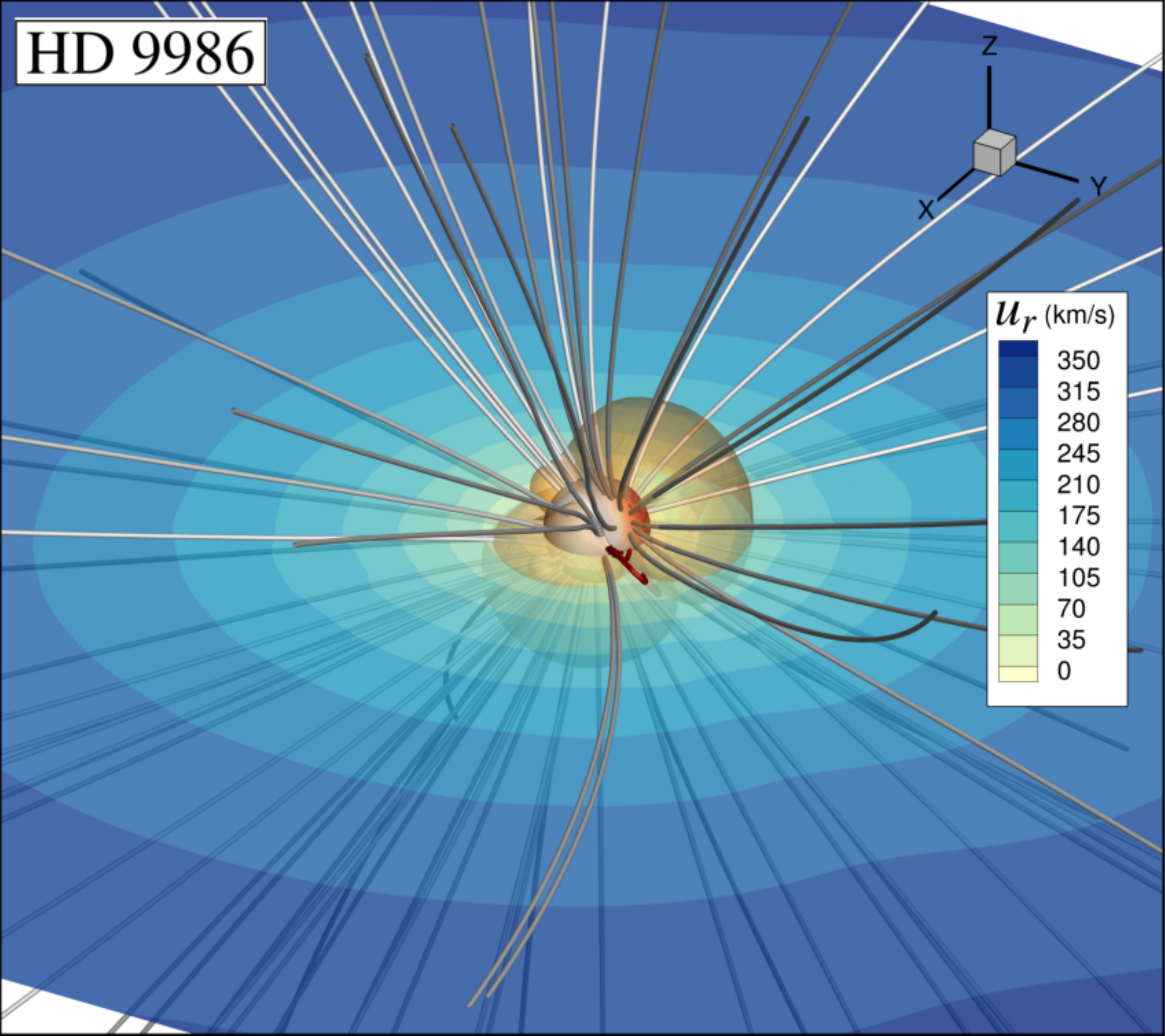}
    \label{fig:hd9986}
    \end{subfigure}

    \begin{subfigure}[b]{.43\linewidth}
    \includegraphics[width=\linewidth]{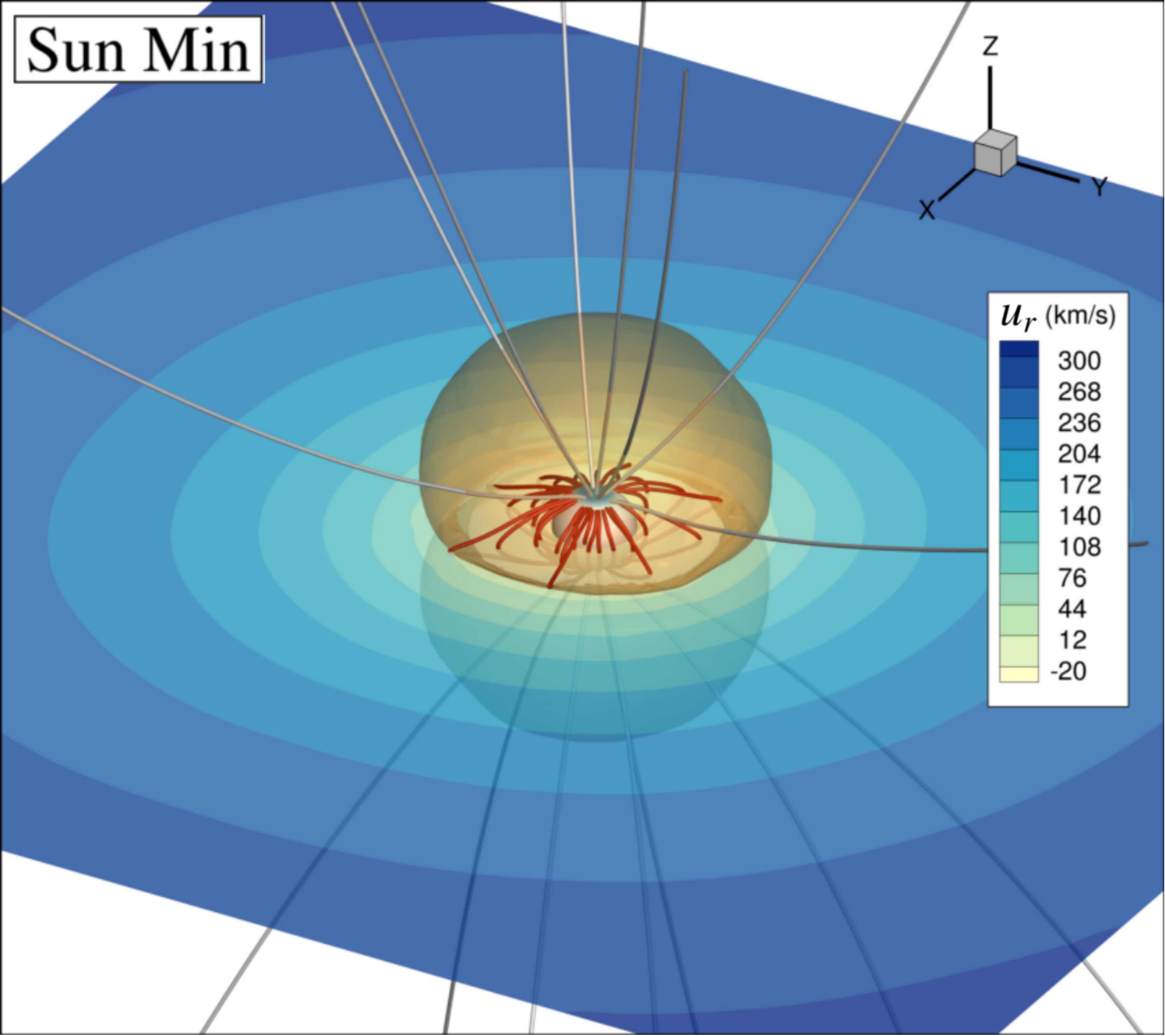}
    \label{fig:sunmin}
    \end{subfigure}   
    \begin{subfigure}[b]{.43\linewidth}
    \includegraphics[width=\linewidth]{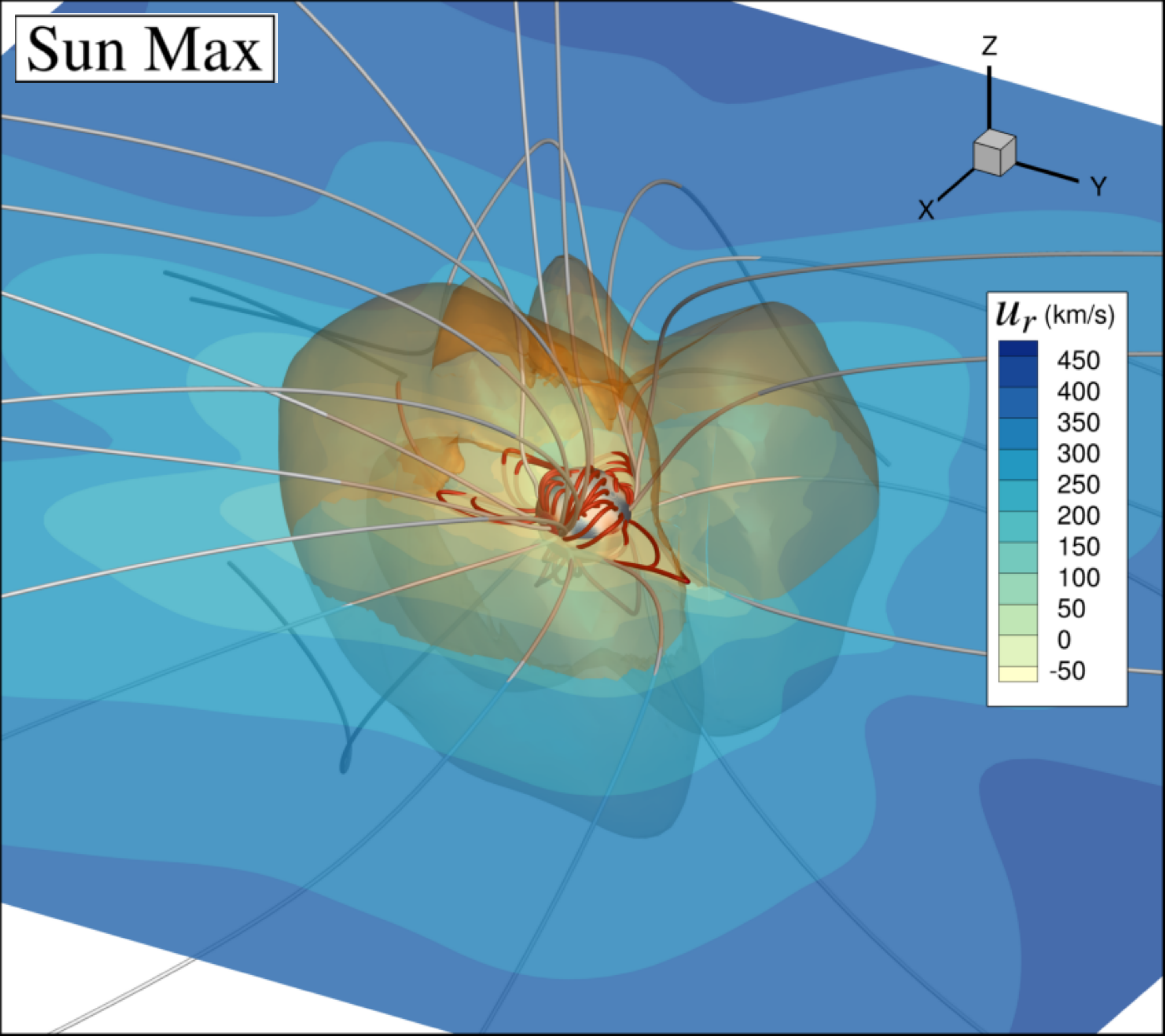}
    \label{fig:sunmax}
    \end{subfigure}
    \caption{(cont.) Steady state solutions for the simulated winds of the solar analogues, showing the slower rotators in our sample.}
    \label{fig:tecplot2}
\end{figure*}
\section{Wind Modelling}
\label{sec:modelling}
 \subsection{3D numerical simulations of stellar winds} 
 We use the 3D MHD numerical code BATS-R-US to simulate the winds of our sample of stars. This code has been used frequently in the past to study many magnetic astrophysical plasma environments \citep{Powell1999,Toth2005,ManchesterIV2008a,Vidotto2015,Vidotto2017,Alvarado-Gomez2018}. Here we use it to solve for 8 parameters: mass density ($\rho$), wind velocity ($\textbf{u} = \lbrace u_x, u_y, u_z \rbrace$), magnetic field ($\textbf{B} = \lbrace B_x, B_y, B_z \rbrace $), and gas pressure P. The code numerically solves a set of closed ideal MHD equations representing, respectively, the mass conservation, momentum conservation, the induction equation, and the energy equation:
 \begin{equation}
 \frac{\partial \rho}{\partial t} + \nabla \cdot (\rho \textbf{u}) = 0,
 \end{equation}
 \begin{equation}
 \frac{\partial (\rho \textbf{u})}{\partial t} + \nabla \cdot \left[ \rho \textbf{uu} + \left( P + \frac{B^2}{8\pi} \right)I - \frac{\textbf{BB}}{4\pi} \right] = \rho \textbf{g},
 \end{equation}
 \begin{equation}
 \frac{\partial \textbf{B}}{\partial t} + \nabla \cdot (\textbf{uB} - \textbf{Bu}) = 0
 \end{equation}
 \begin{equation}
 \frac{\partial \varepsilon}{\partial t} + \nabla \cdot \left[ \textbf{u} \left(  \varepsilon + P + \frac{B^2}{8\pi} \right) -\frac{(\textbf{u} \cdot \textbf{B}) \textbf{B} }{4\pi} \right] = \rho \textbf{g} \cdot \textbf{u},
 \end{equation}
where the total energy density is given by:
\begin{equation}
 \varepsilon = \frac{\rho u}{2} + \frac{P}{\gamma - 1} + \frac{B^2}{8\pi}
 \label{eq:state}
\end{equation}
Here, $I$ denotes the identity matrix, and \textbf{g} the gravitational acceleration. We assume that the plasma behaves as an ideal gas, that $P = n k_B T $, where $n = \rho / (\mu m_p)$ is the total number density of the wind, $\rho$ representing the mass density and $\mu m_p$ denoting the average particle mass. We take $\mu = 0.5$, which represents a fully ionised hydrogen wind. \comment{We can also relate the pressure to the density, by assuming the wind is polytropic in nature, which follows the relationship: $P \propto \rho^{\gamma}$, where $\gamma$ represents the polytropic index.} This polytropic index implicitly adds heat to the wind as it expands, meaning we do not require an explicit heating equation in our model. We adopt $\gamma = 1.05$, which is similar to effective index found by \citet{VanDoorsselaere2011a} for the Sun, and to values used in the literature for simulating winds \citep{Vidotto2015,Pantolmos2017a,OFionnagain2018}.

The free parameters of polytropic wind models, such as ours, are the base density ($\rho_0$) and temperature ($T_0$) of the wind. Here, we use the empirical model from \citet{OFionnagain2018} that relates both the temperature and density of the wind base with the rotation of the star (see also \citealt{Holzwarth2007, See2014, Reville2016, Johnstone2015a, Johnstone2015b}).
    \begin{equation}
    \label{eq:rot-T}
    T_{0}\ (\Omega < 1.4\ \Omega_{\odot}) = 1.5\pm0.19 \left( \frac{\Omega_{\star}}{\Omega_{\odot}} \right) ^{1.2\pm0.54} \rm MK
    \end{equation}
    \begin{equation}
    \label{eq:rot-T2}
    T_{0}\ (\Omega > 1.4\ \Omega_{\odot}) = 1.98\pm0.21 \left( \frac{\Omega_{\star}}{\Omega_{\odot}} \right) ^{0.37\pm0.06} \rm MK
    \end{equation}
    \begin{equation}
    \label{eq:rot-rho}	
    n_0 = 6.72\times10^8 \left( \frac{\Omega_{\star}}{\Omega_{\odot}} \right) ^ {0.6} cm^{-3}.
    \end{equation}
To set the magnetic field vector, we use the radial component of the ZDI maps at the stellar surfaces (\Cref{fig:bfields}). At the initial state, we use a potential field source surface model (e.g. \citealt{Altschuler1969}) to extrapolate the magnetic field into the grid, with the field lines becoming purely radial beyond 4 R$_{\star}$. The code then numerically solves the MHD equations and allows the magnetic field to interact with the wind (and vice-versa), until it reaches a relaxed state. 

\Cref{fig:tecplot1} shows the structure of the winds, with open magnetic field lines displayed in grey and closed magnetic fields shown in red. We can see the field lines become much more structured and organised in the slower rotators with more dipolar fields, as opposed to the complex field lines of the faster rotators with less dipolar fields. Equatorial radial velocities are shown as a yellow-blue graded surface, with the radial velocities ranging from 300-580 km/s at 0.1 au, near the outer boundary of our simulations. Shown in orange are the Alfv\'{e}n surfaces, which denote where the poloidal wind velocity equals the Alfv\'{e}n velocity ($u_{\rm pol} = u_A = B / \sqrt{4\pi \rho}$). They display where the wind becomes less magnetically dominated and more kinetically dominated by the flowing wind. We see these Alfv\'{e}n surfaces range from 2-6 R$_{\star}$ across our sample. Stars with very weak magnetic fields (e.g. 18 Sco, HD 9986) generally have smaller Alfv\'{e}n surface radii.

\subsection[Mass-loss rates, angular momentum-loss rates \& open flux]{Mass-loss rates ($\dot{M}$), angular momentum-loss rates ($\dot{J}$) \& open magnetic flux ($\Phi_{\rm open}$)}
\begin{figure}
    \centering
    \includegraphics[width=\linewidth]{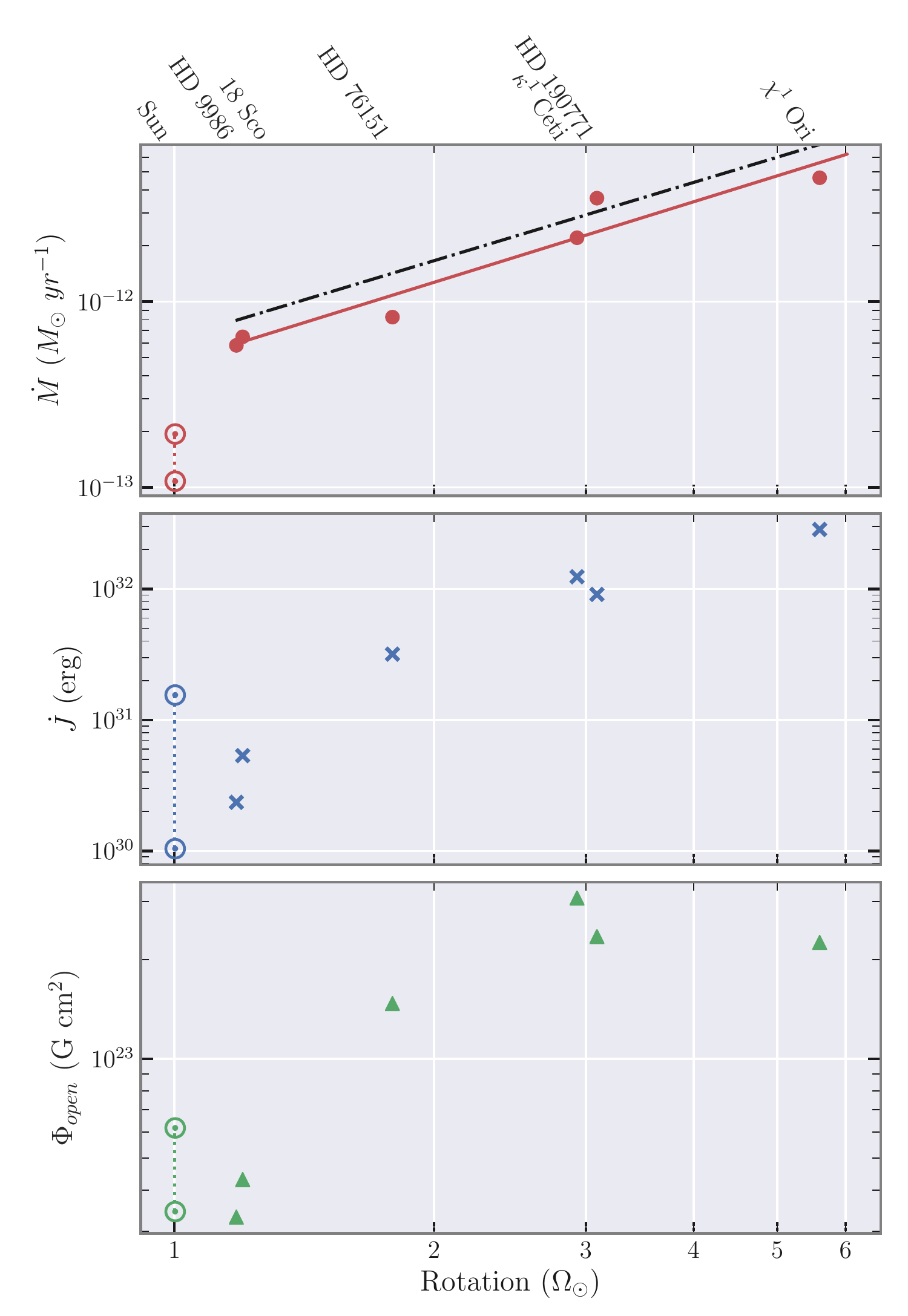}
    \caption{Top to bottom, the three panels above show the mass-loss rate, angular momentum-loss rate, and unsigned magnetic open flux from our sample of simulations. The stars are labelled at the top of the figure, with the solar simulations represented by the solar symbol ($\odot$), where activity maximum is always on top. In the top panel we include a fit to the data (red line, excluding the Sun) and compare this to the fast rotator fit as described in \citet{OFionnagain2018} (black dashed line). }
    \label{fig:global}
\end{figure}

From our wind simulations we can calculate the mass-loss rate from each of the stars by integrating the mass flux through a spherical surface $S$ around the star
\begin{equation}
    \dot{M} = \oint_S \rho u_r dS, 
    \label{eq:mdot}
\end{equation}
where $\dot{M}$ is the mass loss rate, $\rho$ is the wind density, $u_r$ is the radial velocity and S is our integration surface. In our simulations we see an overall decrease of $\dot{M}$ with decreasing rotation rate, \Cref{tab:general}, which is consistent with the works of \citet{Cranmer2011,Suzuki2013a,Johnstone2015a,Johnstone2015b, OFionnagain2018}. We note that the mass-loss rate we find for the Sun is $\approx$ 5 times larger than the observed value of $\sim  2\times10^{-14}$ M$_{\odot}$ yr$^{-1}$. This is because of our choice of base density, which is 3 times higher than in \citet{OFionnagain2018}. We opted for a 3 times higher base density as we were unable to find a stable solution for the winds of a few stars in our sample. 
\citet{OFionnagain2018} suggested that the angular-momentum loss for solar-type stars would drop off substantially for slow rotators, causing older solar-type stars to rotate faster than expected. This would explain the findings of \citet{VanSaders2016}, who observed a set of ageing solar-like stars and discovered that they rotated at much faster rates than expected by the traditional Skumanich age-rotation relationship. In our previous work, \citet{OFionnagain2018}, we linked the anomalous fast rotation at older ages to the drop in mass-loss rates at older ages, and consequently, to a drop in the angular momentum-loss rate. Unfortunately, we could not verify this drop in angular momentum for slower rotators, as we do not have magnetic field maps for solar-mass stars that rotate much slower than the Sun. \dualta{This lack in magnetic field maps in this regime can be explained observationally as detecting weak magnetic fields in slowly rotating stars is very challenging.} Therefore, we compare mass-loss rates calculated here using the faster rotators. \Cref{fig:global} shows the mass-loss rate (red points) and the fit to these points (red line) which follows the relationship
\begin{equation}
    \dot{M} = 4.7(\pm 0.1)\times10^{-13}\ \left(\frac{\Omega_{\star}}{\Omega_{\odot}}\right)^{1.4\pm0.2} M_{\odot} yr^{-1}.
\end{equation}
The fit to the faster rotators from \citet{OFionnagain2018} (shown as a dotted black line), which possesses the power law index of 1.4, agrees within the error to the power law index fit here of 1.6 $\pm$ 0.2. It is interesting that these mass-loss rates agree so well considering the base density of the 3D simulations is 3 times higher than in \citet{OFionnagain2018}. This suggests that the inclusion of a magnetic field in the 3D simulations would generate a much lower mass-loss rate than in the 1D simulations, given the same base densities. This is most likely due to closed magnetic regions, which act to hold in material, and reduce \mdot.

We also determine $\dot{J}$ from our simulations as
\begin{equation}
    \dot{J} = \oint_S \left[ -\frac{\varpi B_{\phi} B_r}{4\pi} + \varpi u_{\phi} \rho u_r \right] dS
\end{equation}
where $\varpi = (x^2+y^2)^{1/2}$, the cylindrical radius,  B and u are the magnetic field and velocity components of the wind, and r and $\phi$ denote the radial and azimuthal components respectively \citep{Mestel1999,Vidotto2014b}. The integral is performed over a spherical surface (S) in a region of open field lines. From \Cref{fig:global} we see a trend of decreasing $\dot{J}$ towards slower rotating stars. We note that while the solar minimum simulation has a reasonable angular momentum loss rate, we find that the solar maximum simulation has a higher $\dot{J}$ than expected (see e.g. \citealt{Finley2018}).

The magnetic field geometry and strength affect the wind in these simulations as it evolves, by establishing a pressure and tension against the ionised plasma. Here we calculate how much of the wind consists of open and closed field lines, by integrating the unsigned magnetic flux passing through a surface near the outer edge of our simulation domain, where all the field lines are open
\begin{equation}
    \Phi_{\rm open} = \oint_{\rm S_{sph}} |B_r|\ dS .
\end{equation}
The open flux of the wind, $\Phi_{\rm open}$, is relevant as regions of open flux the origin of the fast solar/stellar wind \citep{Verdini2010,Reville2016,Cranmer2017}. It is also related to how efficient the wind is at transporting angular momentum from the star \citep{Reville2015}. In \Cref{fig:global} we see that across the rotation periods of our sample, open flux decreases as the stars spin down. There is also a hint of an open flux plateau in the faster rotators. 
In \Cref{tab:general}, we also present the ratio $f$ of open to unsigned surface magnetic field flux ($\Phi_{\rm surf}$), following the convention: $\Phi _{\rm surf} = f \Phi_{\rm open}$. 
\subsection{Wind derived properties at typical hot-Jupiter distances}
 \begin{figure}
     \centering
     \includegraphics[width=\linewidth]{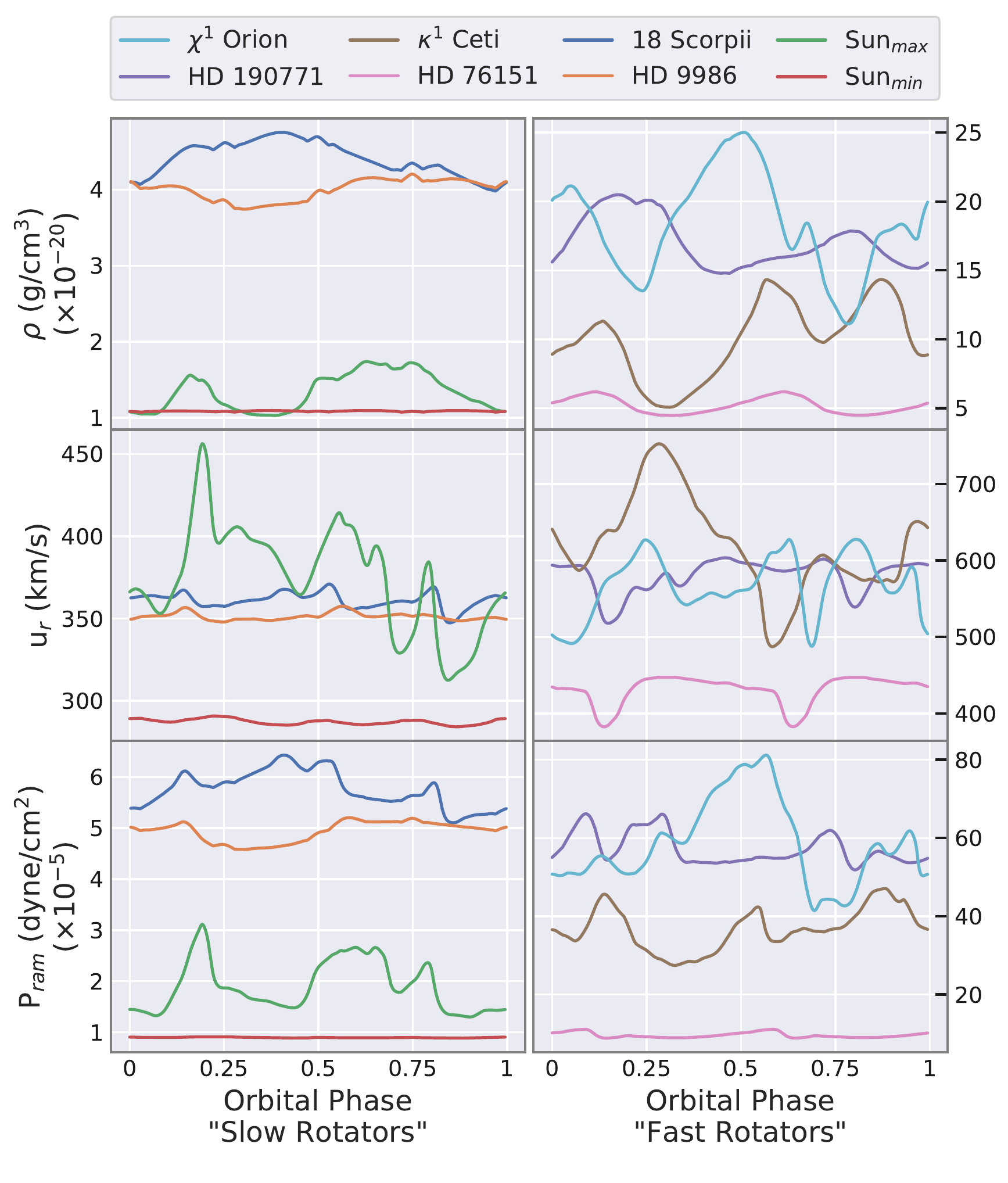}
     \caption{\textit{Top:} Figure showing the density variations of the wind at the equator (z=0 plane) for each star in our sample, at a distance of 0.1 au. \textit{Middle:} The velocity variations of the wind at the equator at 0.1 au. \textit{Bottom:} Calculated ram pressures of the wind at 0.1 au at the equator, using \Cref{eq:ram}. The figures are split into slow (left) and fast (right) rotators so to conserve the visibility of variation across all winds. Note that the y-axes on the left and right have different scales. This figure is optimally viewed in colour.}
     \label{fig:ram}
 \end{figure}
From our simulations we can gather much information on the structure of the winds of solar-like stars. This aids us in the analysis of the wind evolution from young to older solar-type stars along the main sequence. It also impacts the study of exoplanet evolution, as exoplanets exist orbiting these stars, embedded in the stellar wind. The main components of the wind affecting exoplanets are magnetic pressure (for close in exoplanets) and ram pressure (for distantly orbiting exoplanets). There also exists a thermal pressure constituent to the wind, but this is usually much smaller than both of the previous pressures. In our case, at 0.1 au the ram pressure dominates as this is well above the Alfv\'{e}n surface for each star. The ram pressure is given as
 \begin{equation}
     P_{\rm ram} = \rho u_r^2.
     \label{eq:ram}
 \end{equation}
Here we assume the orbit to be in the equatorial plane aligned with the rotation axis, \dualta{but we note that this might not always be the case for hot Jupiters \citep{Huber2013,Anderson2015}.} We see from \Cref{fig:ram} that there can be large variations in the ram pressure impinging upon an orbiting exoplanet at 0.1 au, both within a single orbit around a particular host star, and between each host star. From these, we infer the evolution of the planetary environment around a solar-like star as it evolves. We see that the Sun at minimum possesses the lowest ram pressure of any of the stars in our sample.
 \begin{figure}
     \centering
     \includegraphics[width=\linewidth]{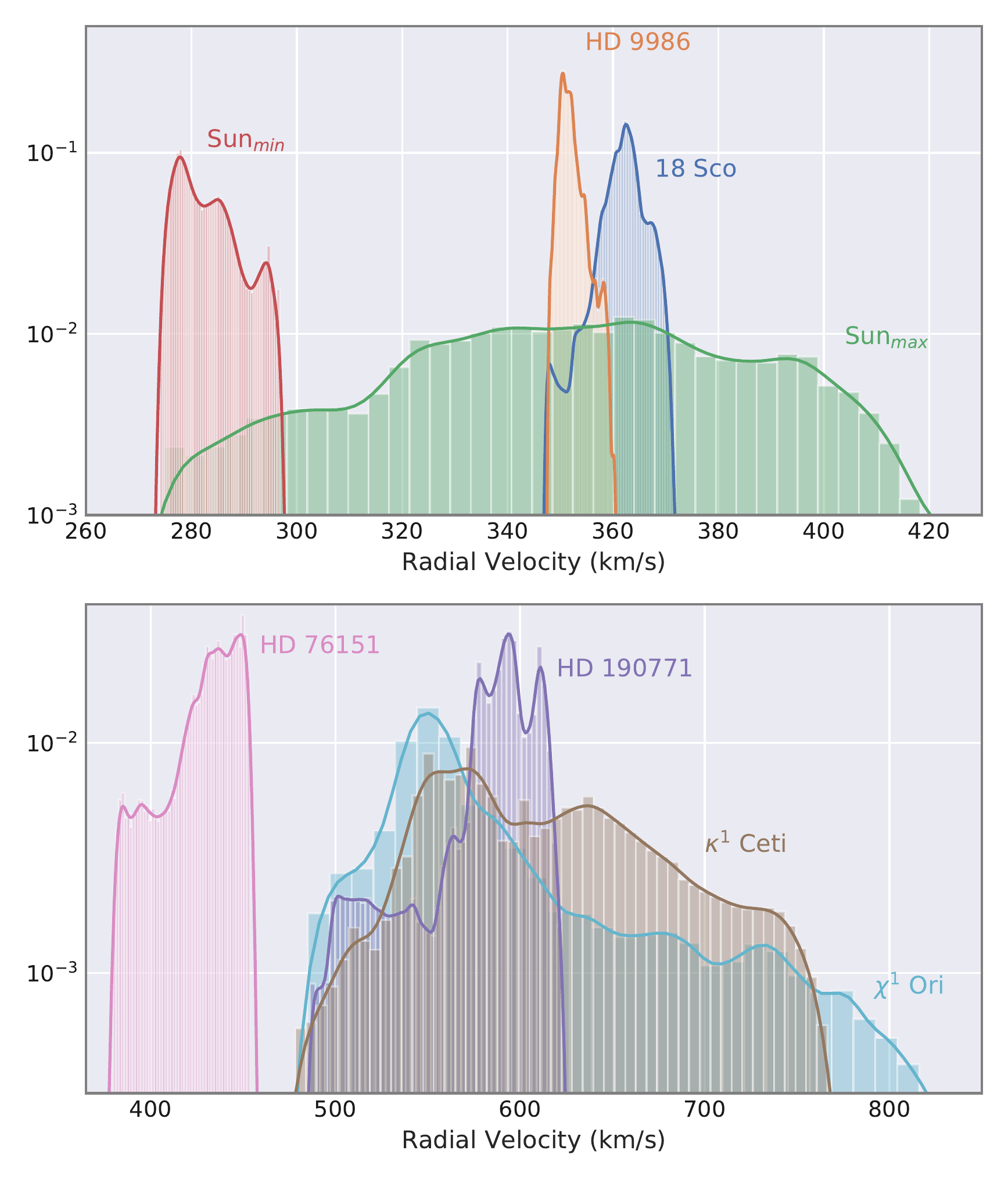}
     \caption{Velocity histogram for our stellar sample, allowing insight into the wind structure (e.g. \citealt{Reville2016}). Velocities are taken at a distance of 0.1 au, \dualta{and split into slower rotators (top) and faster rotators (bottom). Note the different velocity scales on each panel.} This histogram shows the normalised frequency of each velocity present in the wind at this distance. We can see that the winds of 18 Sco and HD 9986 are extremely uni-modal, while other stars such as HD 190771 have a very skewed distribution of velocities. The magnetic field strength and geometry seems to directly affect the wind structure even at these distances. Bin size is selected using the Freedman-Diaconis rule.}
     \label{fig:velocities}
 \end{figure}
We can compare the distribution of velocities for all of the stars by histogramming the velocities across a sphere of 0.1 au. This method can give insight into the structure of the wind, discerning uni-modal and multi-modal wind structures (see \Cref{fig:velocities}). We observe that more complex and stronger fields lead to less uniform wind structures. We can see that the winds of 18 Sco, HD 9986 and the Sun at minimum display uni-modality, while other stars such as $\chi^1$ Ori and HD 190771 have a very skewed velocity distributions. The magnetic field strength and geometry seems to directly affect the wind structure even at these distances. This is discussed in \citet{Reville2016}, who noted that the expansion of magnetic flux tubes can cause an acceleration in the wind.
\section{Radio Emission of the solar wind in time}
\label{sec:radio}
\subsection{Radiative transfer model}
It has long been established that the plasma of stellar winds emit at radio wavelengths through thermal free-free processes \citep{Panagia1975,Wright1975,Lim1996}. If this radio emission is observed, it could provide a way to detect the winds of low-mass stars directly, allowing an estimation of the wind density and temperature at that location in the wind. Constraining the density of the wind would allow a much better estimate on the mass-loss rate of the star, and by extension angular-momentum loss rates. \par
Analytical expressions for the radio emission calculation are commonly used in the literature \citep{Panagia1975,Wright1975,Lim1996,Fichtinger2017,Vidotto2017a}. For example, \citet{Panagia1975} assumed a power law dependence of density with radial distance, such that $\rho \propto R^{-\alpha}$, which generates a radius dependence for radio flux density with frequency: $S_{\nu} \propto \nu^{\frac{-4.2}{2\alpha-1} +2}$. However, when R is small and the wind is still accelerating, this density dependence deviates from a power law. Thus, these power law gradients can underestimate the density decay close to the star and overestimate it further from the star. This is discussed further in \Cref{app:density}. A similar approach is also used in defining the distance-dependence of the temperature of the wind. 
To overcome this, we perform the radio emission calculation from first principles, by solving the radiative transfer equation numerically (code available on GitHub: \href{https://github.com/ofionnad/radiowinds}{https://github.com/ofionnad/radiowinds}; \citealt{radiowinds}). Using our 3D MHD simulations, we can use the exact density decay expected, which gives a more precise estimation of the wind emission. 

\begin{figure}
      \centering
      \includegraphics[width=\linewidth]{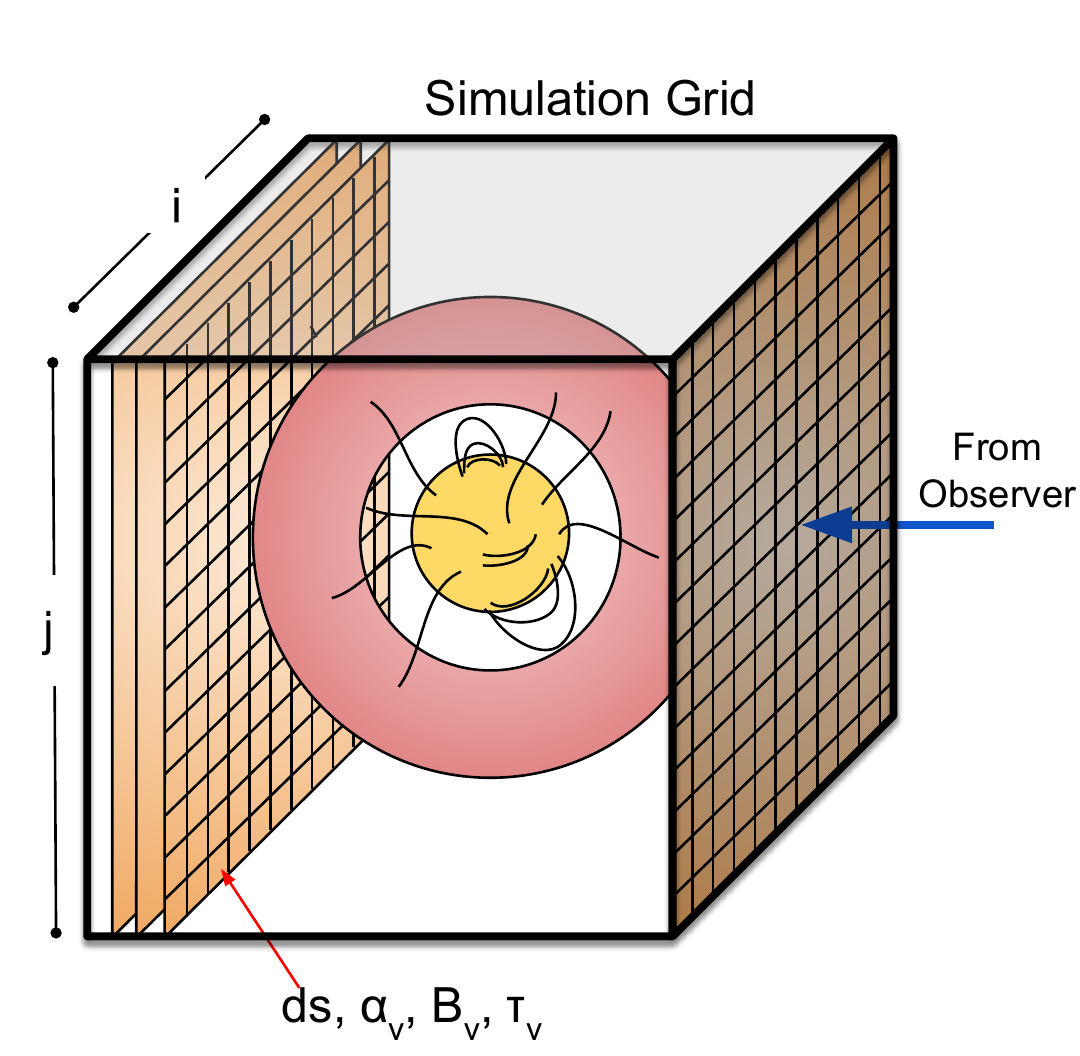}
      \caption{Schematic showing how the intensity is calculated from our grid. The red annulus around the star illustrates thermal radio emission regions from the wind, with a magnetic star at the centre of the diagram. From our wind simulation we create a grid of uniform discrete distances filled with variables including position (s), density (n) and temperature (T). From this we calculate values for the absorption coefficient ($\alpha$, \cref{eq:absorption}), the blackbody function ($B_{\nu}$, \cref{eq:blackbody}) and the optical depth ($\tau_{\nu}$, \cref{eq:tau}) for each cell in our grid. We integrate along the line-of-sight from the observer to find the intensity using \Cref{eq:intensity}, and find flux density by integrating across i and j. We take the line-of-sight to be along the x axis for each star, which is not necessarily true, but adopted as such because it is assumed variability in the radio emission will not vary much depending on viewing angle or rotation axis.}
      \label{fig:integration_sketch}
  \end{figure}

\Cref{fig:integration_sketch} shows a schematic of our calculation grid, we divide the grid into equally spaced cells, each possessing a value of wind density and temperature. The illustration shows a red annulus around a magnetic star, outlining the expected radio emission from the wind (this is not expected to be spherically symmetric). Note that the actual number of cells used in calculations (~=~200$^{3}$) is much greater than depicted in \Cref{fig:integration_sketch}. From this, we can calculate the thermal emission expected from these winds by solving the radiative transfer equation, 
\begin{equation}
    I_{\nu} = \int_{-\infty}^{\tau'_{\rm max}} B_{\nu} e^{-\tau} d\tau'
    \label{eq:intensity}
\end{equation}
where I$_{\nu}$ denotes the intensity from the wind, B$_{\nu}$ represents the source function, which in the thermal case becomes a blackbody function, $\tau$ represents the optical depth of the wind, with $\tau'$ representing our integration coordinate across the grid. The optical depth of the wind depends on the absorption coefficient, $\alpha_{\nu}$, of the wind as 
\begin{equation}
	\tau_{\nu} = \int \alpha_{\nu} ds, 
	\label{eq:tau}
\end{equation}
where \textit{s} represents the physical coordinate along the line of sight, $\alpha_{\nu}$ is described as \citep{Panagia1975,Wright1975,Cox2000}, 
\begin{equation}
    \alpha_{\nu} = 3.692\times10^8 [1 - e^{-h\nu / k_BT}] Z^2 f_g T^{-0.5}\nu^{-3} n_e n_i
    \label{eq:absorption}
\end{equation}
and the blackbody function is the standard Planck function.
\begin{equation}
    B_{\nu} = \frac{2h\nu^3}{c^2} \frac{1}{e^{hv/k_B T} -1}
    \label{eq:blackbody}
\end{equation}
where $\nu$ is the observing frequency, h is Planck's constant, k$_B$ is Boltzmann's constant, T is the temperature of the wind, Z is the ionic state of the wind (+1 for our ionised hydrogen wind), with n$_e$ and n$_i$ representing the electron and ion number densities of the wind. In our case we have the same number of ions and electrons, so this becomes simply n$_i^2$. \dualta{f$_g$ is the gaunt factor} which is defined as \citep{Cox2000}
\begin{equation}
    f_g = 10.6 + 1.9 \log_{10} T - 1.26 \log_{10} Z\nu
\end{equation}
\par
  \begin{figure*}
    \centering
    \begin{subfigure}[b]{.43\linewidth}
    \includegraphics[width=\linewidth]{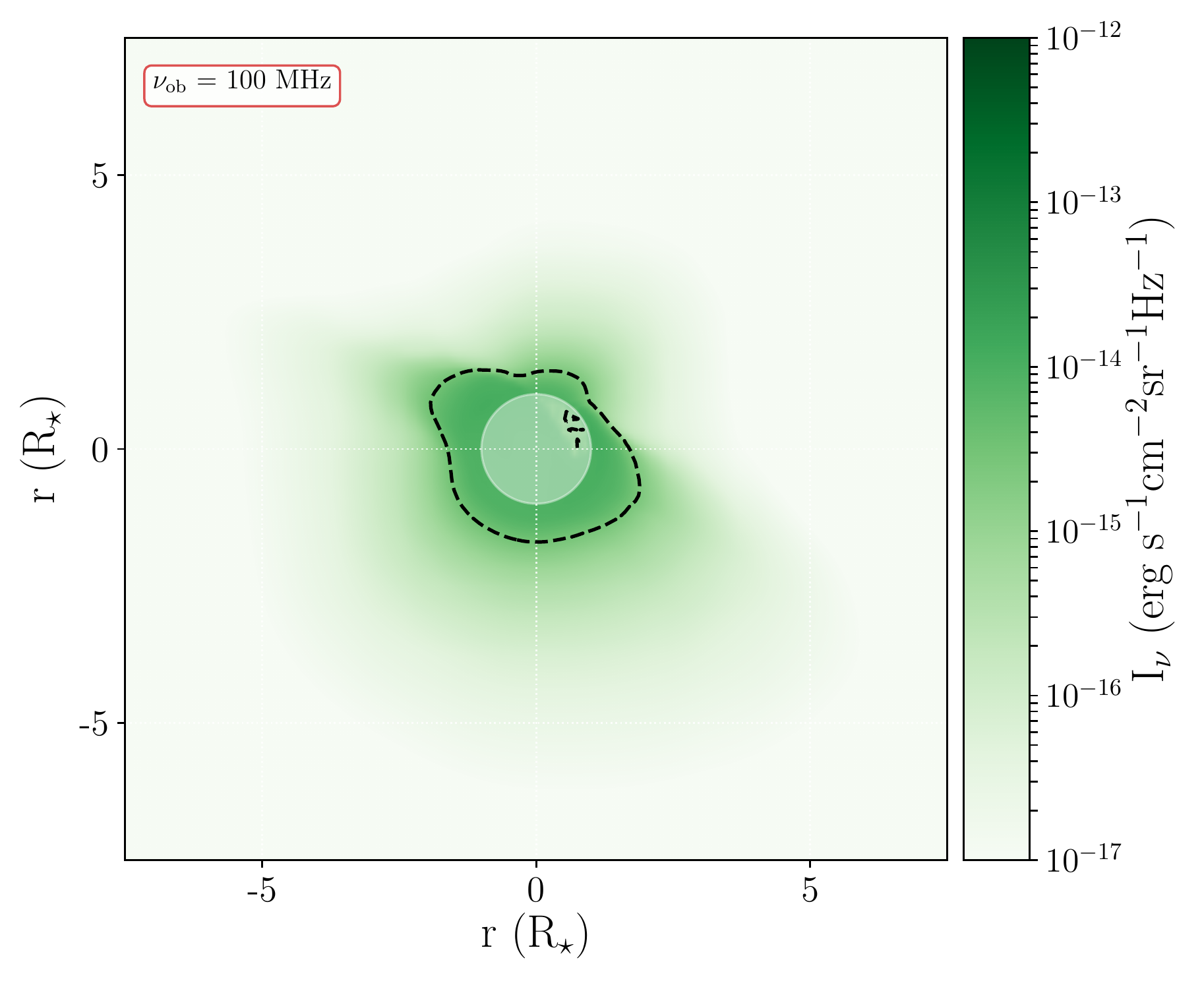}
    \label{fig:100mhz}
    \end{subfigure}
    \begin{subfigure}[b]{.43\linewidth}
    \includegraphics[width=\linewidth]{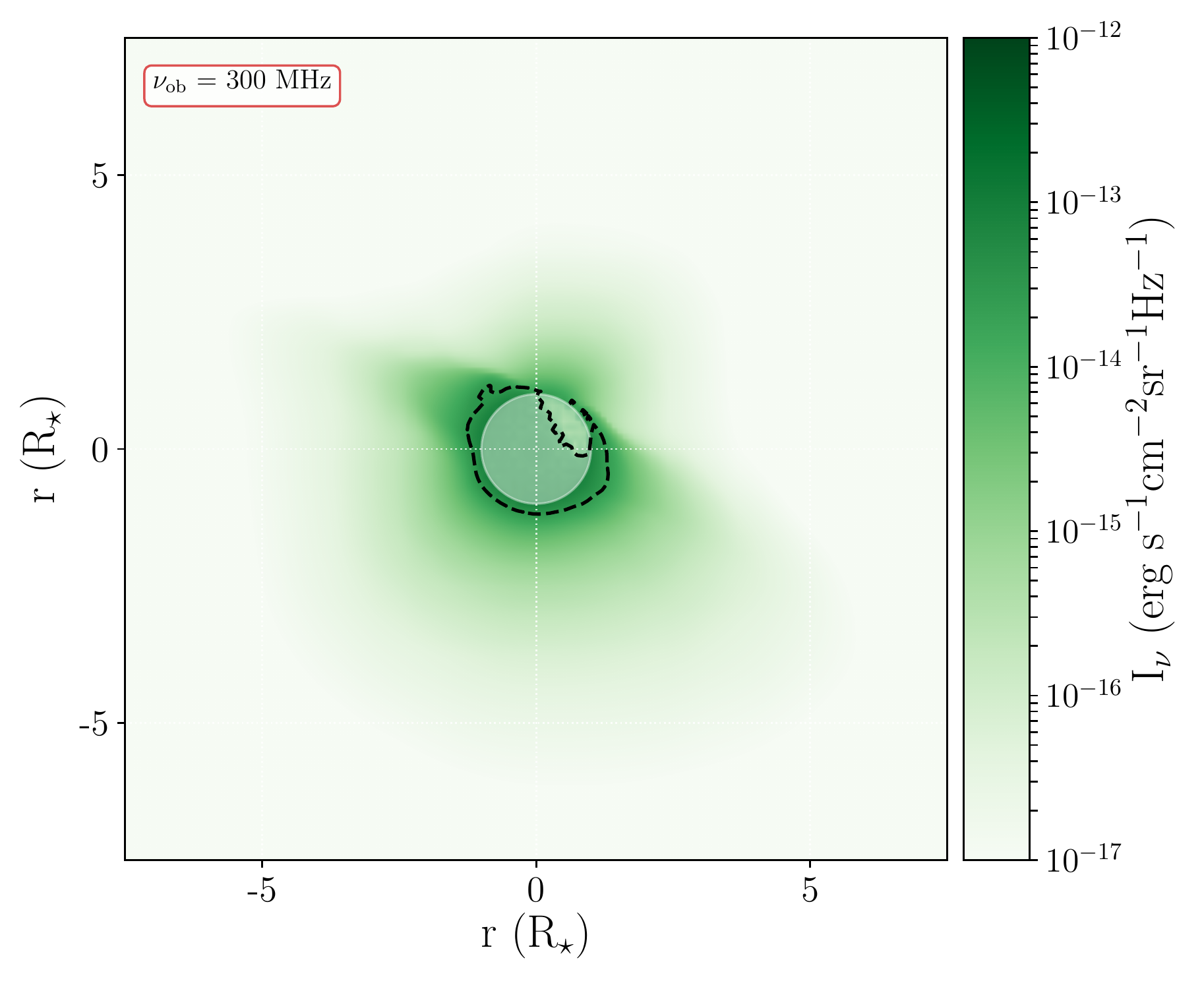}
    \label{fig:300mhz}
    \end{subfigure}
    
    \begin{subfigure}[b]{.43\linewidth}
    \includegraphics[width=\linewidth]{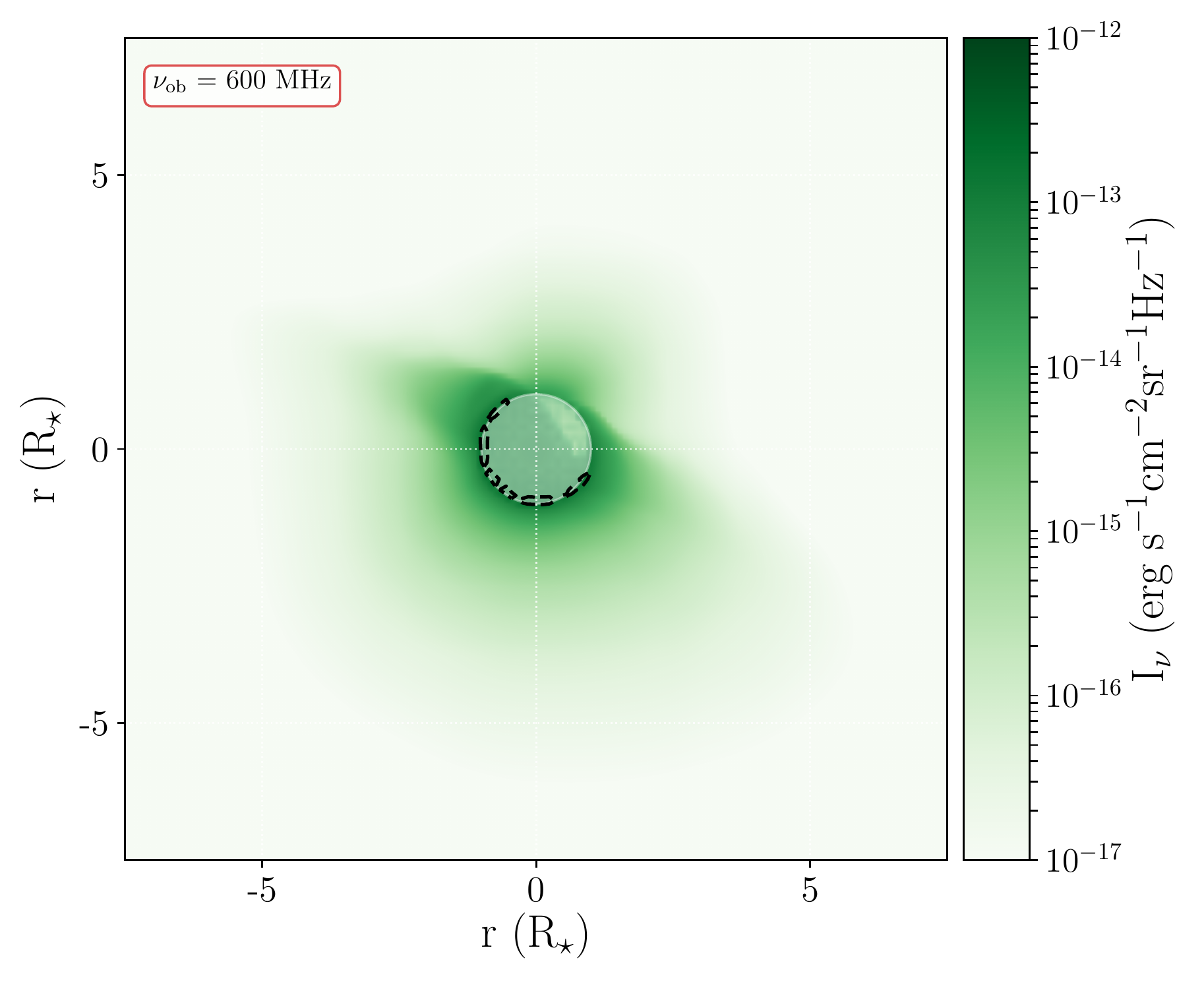}
    \label{fig:1ghz}
    \end{subfigure}
    \begin{subfigure}[b]{.43\linewidth}
    \includegraphics[width=\linewidth]{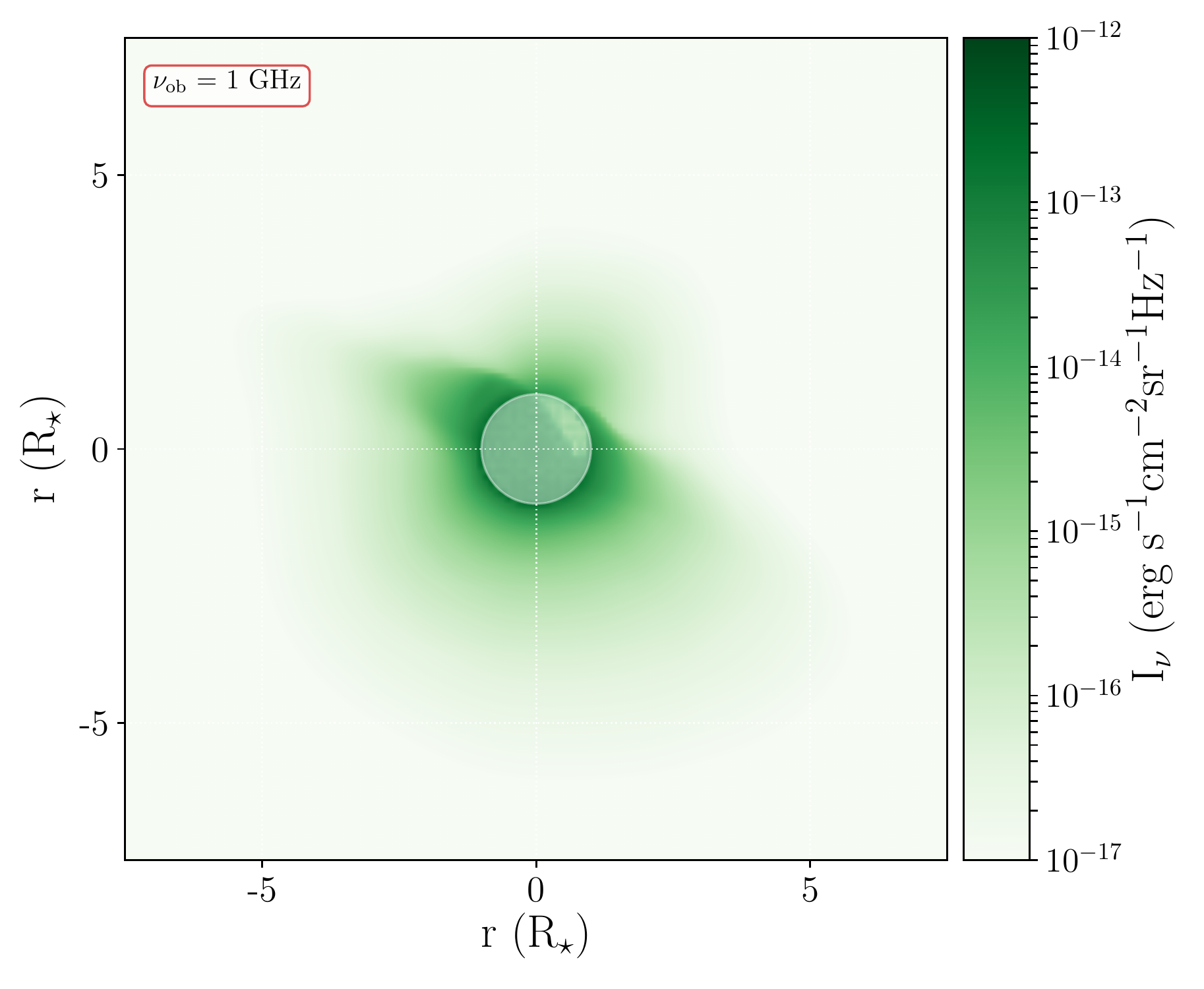}
    \label{fig:6ghz}
    \end{subfigure}
    \caption{Example of intensity and optical depth for $\kappa^1$ Ceti at observing frequencies of 100 MHz (top left), 300 MHz (top right), 600 MHz (bottom left) and 1 GHz (bottom right). The green colour scale represents the intensity of emission from the wind, looking along the line-of-sight of our simulation grid. The dashed black contour represents the region where the wind becomes optically thick, according to \citet{Panagia1975}, $\tau$ = 0.39). We can see that the emission is anisotropic due to the anisotropy of the wind density and temperature. The intensity reaches a maximum in the thin regime, as we can see emission from the entire wind. The white circle denotes R = 1 R$_{\star}$. Plasma in front of the star still emits in radio, but we have excluded any contribution from behind the star along the line-of-sight.}
    \label{fig:radio_intensity}
\end{figure*}
\subsection{Evolution of the radio emission with age}
\label{sec:radio-age}
Using \Cref{eq:intensity,eq:tau,eq:absorption,eq:blackbody} we calculate 2D images for each frequency (cube of data) for the intensity and optical depth, across the plane of the sky, showing the intensity attributed to different regions of the wind, and the optical depth associated with it. This is represented in \Cref{fig:radio_intensity}. \dualta{Note that for comparison we calculate solar wind radio emission at a distance of 10 pc.} We can see that the intensity of the emission increases as we increase the frequency, although it radiates from a much smaller region. This is due to the decrease in the optical depth with frequency and allows us to see further into the wind, to much denser regions giving rise to more emission.  The optical depth of the wind will have a major impact on the observations of these winds. Low optical depths allow emission from the low corona to escape and be detected, these regions are contaminated with other forms of radio emission, likely dominant, such as chromospheric emission and flaring. However, \citet{Lim1996} suggest that we still can provide meaningful upper limits to the mass-loss rate of the star if a flare is detected as one must assume a maximum base density to the wind, therefore constraining mass-loss rates.
From the intensity we can calculate the flux density (S$_{\nu}$) of the wind as, 
\begin{equation}
    S_{\nu} = \frac{1}{d^2} \int I_{\nu} dA = \frac{1}{d^2} \sum^{i,j} I_{\nu}\ \Delta i\ \Delta j
\end{equation}
where $A$ is the area of integration, d is the distance to the object, and i and j denote the coordinates in our 2D image of I$_{\nu}$ values. $\Delta$i and $\Delta$j represent the spacing in our grid in the i and j directions. \comment{In this calculation we have assumed that the angle subtended by the stellar wind is small, therefore d$\Omega = dA / d^{2}$.}

\begin{figure}
 	\centering
 	\begin{subfigure}[b]{\linewidth}
     \includegraphics[width=\linewidth]{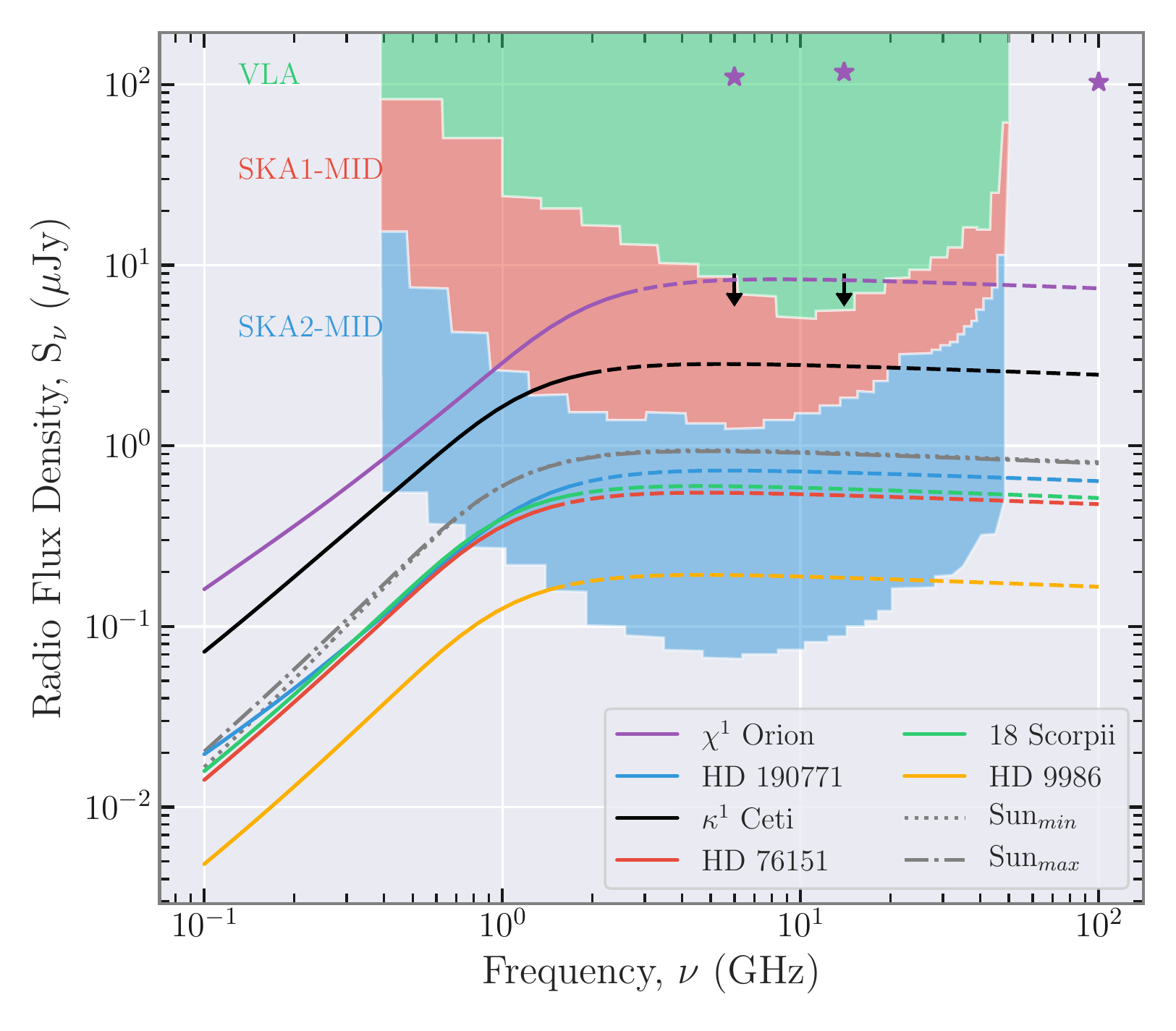}
      \caption{}
      \label{fig:spectra}
    \end{subfigure}
    \begin{subfigure}[b]{\linewidth}
     \includegraphics[width=\linewidth]{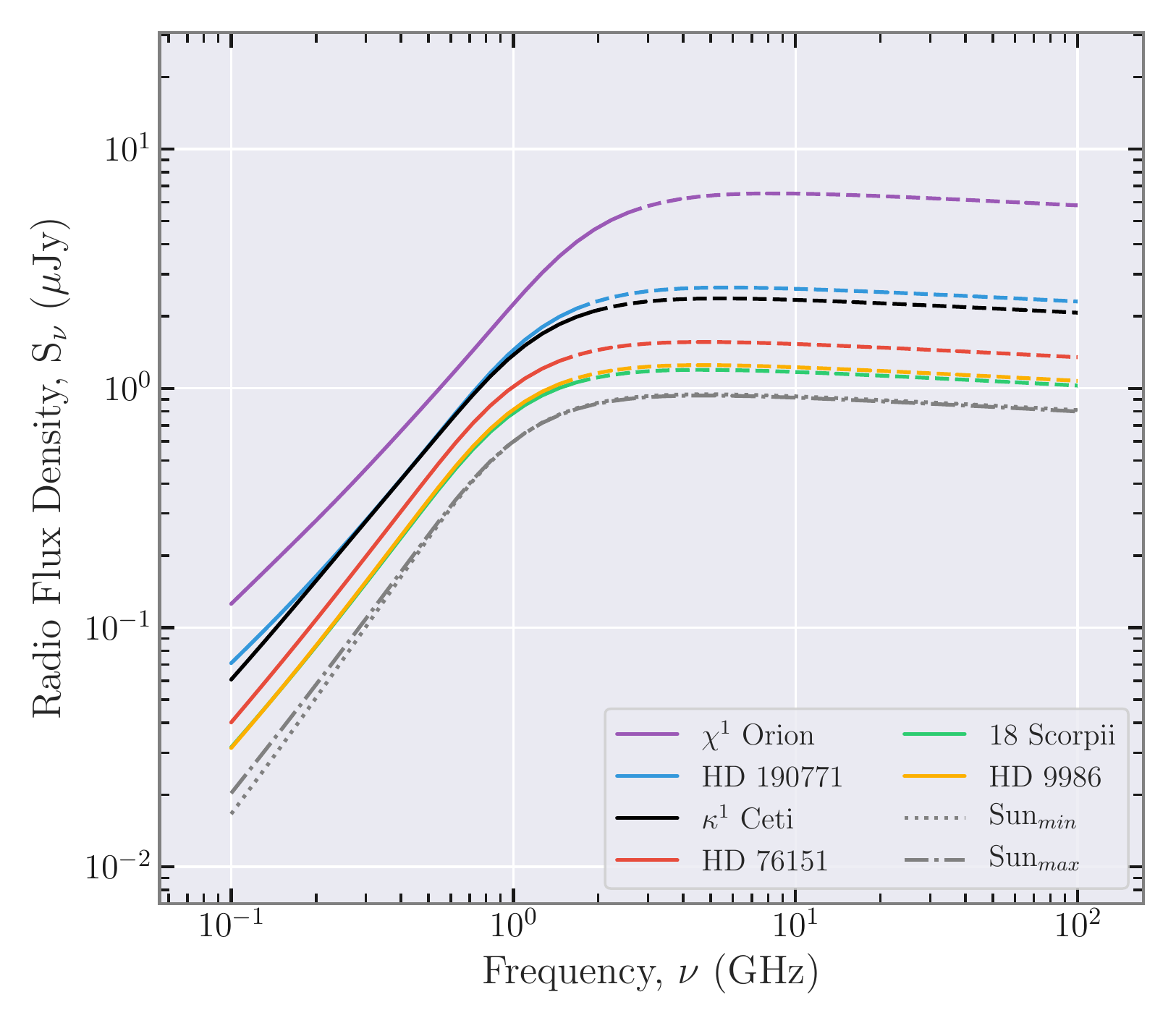}
     \caption{}
     \label{fig:norm_spectra}
    \end{subfigure}
    \caption{\emph{Top:} We see that the radio spectra for each wind are very similar in shape. Differences in flux density are strongly affected by distance to the object. The dashed lines represent the optically thin part of each spectrum, and there are differences in where the emission becomes optically thin from star to star at the frequency $\nu_{\rm thin}$. The black arrows indicate the observational upper limits of $\kappa^1$ Ceti found by \citet{Fichtinger2017}. From the same work we mark the chromospheric detections of $\chi^1$ Ori (purple stars), using both VLA and ALMA, which is concluded to originate from chromospheric emission. Our results show this conclusion to be valid as we predict the wind to emit at much lower fluxes. Sensitivities of the current VLA and future SKA1-MID and SKA2-MID are included shaded in green, red, and blue respectively (SKA sensitivities from \citealt{Pope2018} and adjusted for 2 hour integration time. \emph{Bottom:} Here we normalised spectra in the top panel to a distance of 10 pc. This allows direct comparison of radio emission to an ageing solar wind. As the stars age and spin down the radio emission decreases by an order of magnitude between 500 Myr and 4.6 Gyr.}
 \end{figure}

\Cref{tab:radio-emission} shows the main results from our radio emission calculation, giving values for the expected flux density, from each star at 6 GHz. \Cref{fig:spectra} shows the spectrum of each stellar wind for the range of frequencies 0.1-100 GHz. Our calculation uses actual density distribution in the simulated wind to find the optical depth and the flux density. \dualta{We obtain a spectrum in the optically thick regime, leading to a power law fit which is related to the density gradient in the wind.} Another result of using a numerical model is that the radio photosphere (R$_{\nu}$), calculated at a distance where $\tau = 0.399$, is not spherical, but changes with the density variations in the wind, causing anisotropic emission, as evident from \Cref{fig:radio_intensity} (dashed contours). \comment{Note that these radio winds are not resolvable with current radio telescopes but should indicate how the radio photosphere in the wind changes with frequency, and the anisotropy of the specific intensity, I$_{\nu}$, in the wind.} We also provide a power law fit to the optically thick regime of the radio emission (from 0.1-1 GHz) and note that it can vary quite significantly, depending on what range of frequencies is being fitted. \dualta{In \Cref{tab:radio-emission} we show the fit parameters we find according to }
\begin{equation}
    S_{\nu} = S_0 \nu ^\phi.
    \label{eq:flux-fit}
\end{equation}
  \begin{table}
  \centering
  \caption{Predicted radio emission from our stellar wind models. Example fluxes at a frequency of 6 GHz are given (S$_{\rm 6GHz}$), in this case we find that all of the winds would be optically thin at this frequency. The power law fit to the spectra was conducted between 0.1 and 1 GHz, giving the coefficient (S$_0$) and power index ($\phi$). However, the spectral slope between these two frequencies varies substantially, tending to shallower slopes at higher frequencies. Depending on the fitting range, slopes can range from 0.6 to 1.5. All slopes tend to $-0.1$ in the thin regime. The final column gives the frequency at which each wind becomes optically thin ($\nu_{\rm thin}$).}
    \begin{tabular}{lcccc}
  \hline
  Star & S$_{\rm 6GHz}$ ($\mu$Jy) & S$_{0}$ & $\phi$ & $\nu_{\rm thin}$ (GHz) \\
  \hline \hline
  $\chi^1$ Ori & 8.28 & 2.78 & 1.26 & 2.80 \\
  HD 190771 & 0.73 & 0.39 & 1.32 & 1.85 \\
  $\kappa^1$ Ceti & 2.83 & 1.67 & 1.35 & 2.13 \\
  HD 76151 & 0.55 & 0.37 & 1.41 & 1.61 \\
  18 Sco & 0.60 & 0.40 & 1.40 & 1.63\\
  HD 9986 & 0.19 & 0.13 & 1.42 & 1.63 \\
  Sun max (10 pc) & 0.94 & 0.63 & 1.55 & 2.01\\
  Sun min (10 pc) & 0.93 & 0.62 & 1.47 & 2.00\\
  \hline
  \end{tabular}
  \label{tab:radio-emission}
  \end{table}
Our radio calculations give an insight into the expected emissions from solar-type stars. We see that, at the appropriate sensitive frequencies for radio telescopes such as the VLA, the winds all exhibit similar spectrum shapes. \Cref{fig:spectra} shows the spectrum for each star, using different colours as depicted in the legend. We show that the upper limits set by \citet{Fichtinger2017} (hereafter, F17) (black arrows) are consistent with our estimations of the wind emission for $\kappa^1$ Ceti: our values are 3 times lower than these upper limits. \comment{$\chi^1$ Ori is detected by F17, but they attribute this emission to the chromosphere and other sources as the star was observed to flare during the observation epoch (we discuss detection difficulties further in \Cref{sec:detectability}). Indeed, \Cref{fig:spectra} shows the detected emission occurs within the optically thin regime of the spectrum according to our models and at approximately 20 times higher flux density than we predict for the stellar wind emission. This supports the deduction that these detections are from other sources, and not the thermal wind. F17 estimated the thermal wind emission to possess a flux of 1.3 $\mu$Jy at 10 GHz, which agrees quite well with our calculation of 0.77 $\mu$Jy. If the emission seen at 100 $\mu$Jy by F17 were coming from the stellar wind, our models would require a base density 5 times larger ($\approx$ $\rm 10^{10} cm^{-3}$). With this, we can actually infer that the mass-loss rate of $\chi^1$ Ori is smaller than 1.4 $\rm \times10^{-11} M_{\odot}$ yr$^{-1}$, showing that even non-detections of stellar wind radio emission can still provide meaningful upper limits for the mass-loss rates.} If we normalise the spectra shown in \Cref{fig:spectra} to remove the distance dependence, upon which the spectrum relies very heavily, we see that the younger more rapidly rotating stars display a higher flux density than the more evolved stars. The Sun in this case would possess the weakest emission.
\subsection{Evolution with magnetic cycle}
In \Cref{fig:spectra} we calculate the expected radio emission from our solar maximum and solar minimum simulations assuming a distance of 10pc (grey lines) to give an impression of the differences between the radio emission of the winds and the detectability of each star. We show that the thermal quiescent radio flux does not change substantially across a solar magnetic cycle. This is because the radio emission is heavily dependent on the density of the medium and both solar simulations have the same base density. The slight spectral differences, which occur mostly in the optically thick regime, are a consequence of the different magnetic fields causing different density gradients in the wind. For there to be substantial differences in thermal radio emission from a star displaying cyclic magnetic behaviour there would need to be a dramatic change in global density at the base of the wind. \dualta{Note that the emission calculated here is quiescent wind emission and is the same in both the solar maximum and minimum cases. Non-thermal radio emission, such as 10.7 cm emission, is linked to solar activity and varies through the solar activity cycle \citep{Solanki2010}.}
\subsection{Detectability}
\label{sec:detectability}
\begin{figure}
    \centering
    \includegraphics[width=\linewidth]{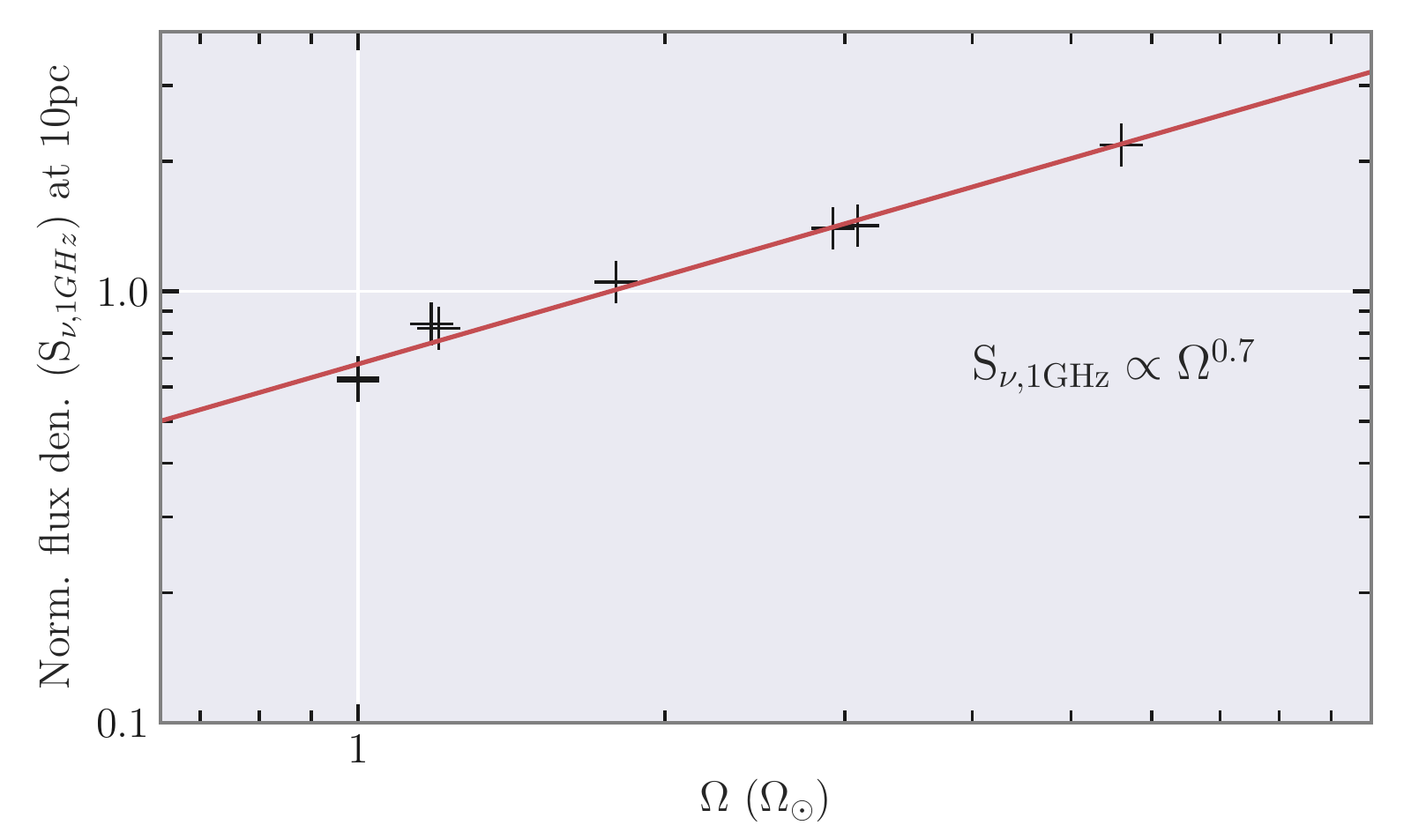}
    \caption{Normalised flux density at 1 GHz as a function of stellar rotation. We see a tight fit to this power-law (see \Cref{eq:flux-rot}), with an almost linear dependence of stellar wind radio flux on stellar rotation at 1 GHz.}
    \label{fig:flux_v_rot}
\end{figure}
\comment{The density at low heights in the stellar atmosphere is much higher than the stellar wind density. Radio emission from the lower atmosphere should dominate the emission in the optically thin regime of the stellar wind. This would most likely drown out any emission from the wind in the upper atmosphere and make detection of the wind impossible. However, as pointed out by \citet{Reynolds1986}, if the wind is entirely optically thin and emission is deduced to emanate from the lower stellar atmosphere, this can aid in placing limits on the stellar winds density and therefore the mass-loss rate of the star (cf. end of \Cref{sec:radio-age}).}\par
There have been many observations of solar-type low-mass stars in the radio regime \citep{Gudel1998, Gaidos2000,Villadsen2014,Fichtinger2017}, many of which have placed upper flux densities and mass-loss rates on the winds of these stars. Both \citet{Gaidos2000} and F17 used the VLA to observe a set of solar analogues, some of which overlap with the stars we have simulated here, placing tight constraints on the wind of $\kappa^1$ Ceti. \Cref{fig:spectra} displays the sensitivity of the VLA (purple shade) given some typical observational parameters (2 hour integration time, 128 MHz bandwidth) taken at central band frequencies. We show that the VLA is currently not sensitive enough to detect the winds simulated here.
\citet{Villadsen2014} observed four nearby solar-like stars using the VLA (X, Ku and Ka bands, at 10 GHz, 15 GHz, and 34.5 GHz centre frequencies respectively). The authors find detections for all objects in the Ka band but can only provide upper limits to flux density for the other frequency bands. They conclude (similarly to F17) that all detections come from thermal chromospheric emission, and the upper limits set at lower frequencies infer rising spectra and so optically thick chromospheres at these frequencies. 

In the future, upgrades to the existing VLA system (ngVLA, see \citealt{Osten2018}) could increase instrument sensitivity by a factor of 10. This increase in sensitivity means that stars simulated here such as $\chi^1$ Ori \& $\kappa^1$ Ceti would be detectable in their thin regime. The Square Kilometre Array (SKA) project is a future low-frequency radio telescope that will span a large frequency range. \comment{The expected sensitivity level of the future SKA1-MID and SKA2-MID telscopes (with a typical 2 hour integration time\footnote{\url{https://astronomers.skatelescope.org/wp-content/uploads/2016/05/SKA-TEL-SKO-0000002_03_SKA1SystemBaselineDesignV2.pdf}}) are shown in \Cref{fig:spectra}, shaded in red and blue (sensitivities for SKA taken from \citealt{Pope2018}, but adjusted to account for a 2 hour integration time). Given these sensitivities one could potentially directly detect the winds of $\chi^1$ Ori \& $\kappa^1$ Ceti using the SKA, below 1 GHz. This sensitivity level (sub-$\mu$Jy) means other possible solar analogues not simulated here could also be detected, provided they are close enough.} First light for SKA1-MID is expected after the mid 2020's. \par
\dualta{We show in \Cref{fig:norm_spectra} that the faster rotators emit more flux. In \Cref{fig:flux_v_rot}, we present the normalised flux density at 1 GHz and at a distance of 10pc as a function of rotation rate. We found that
\begin{equation}
    \label{eq:flux-rot}
    S_{\rm \nu,1GHz} = 0.68 \left[ \frac{\Omega}{\Omega_{\odot}} \right]^{0.7} \left[ \frac{10pc}{d} \right]^2 \rm{\mu Jy} 
\end{equation}
Consequently, younger, rapidly rotating stars within a distance of 10 pc will be the most fruitful when observing thermal radio emission from stellar winds.}

\section{Summary \& Conclusions}
\label{sec:conclusion}
In this study, we presented wind simulations of 8 solar-analogues (including 2 of the Sun itself, from Carrington rotations 1983 and 2078) with a range of rotation rates and ages, using a fully 3D MHD code, (\Cref{fig:tecplot1}). \dualta{We selected a sample of solar-type stars and constrained the sample for which we had observations of their surface magnetic fields (\Cref{fig:bfields}.)} Other input parameters for our model include base temperatures and densities retrieved from semi-empirical laws scaled with rotation, \Cref{eq:rot-T,eq:rot-T2,eq:rot-rho} \citep{OFionnagain2018}.

We demonstrated that the angular-momentum loss rate decreases steadily along with mass-loss rate over evolutionary timescales (\Cref{fig:global}). Younger stars ($\rm \approx 500 Myr$) rotating more rapidly ($\rm P_{\rm rot} \approx$ 5 days) display \jdot\ values up to $\approx10^{32}$ ergs. \dualta{The Sun (4.6 Gyr, $\rm P_{rot} = 27.2$ days) alternatively exhibits a much lower \jdot\ at minimum $\approx10^{30}$ ergs, with significant variance of one order of magnitude over the solar magnetic cycle. The difference in solar \jdot\ from minimum to maximum is explained by the greater amount of $\Phi_{\rm open}$ in the solar maximum case. Given that our solar maximum and minimum simulations differ, this incentivises the monitoring of stars across entire magnetic cycles to deepen our understanding of stellar activity cycles \citep{Jeffers2017,Jeffers2018}.} We found a similar declining rotation trend with \mdot\ , with slower rotators losing less mass than their faster rotating counterparts. Our solar analogues display a \mdot\ ranging from $1\times10^{-13} - 5\times10^{-12} M_{\odot}$ yr$^{-1}$.

We showed in \Cref{fig:ram} how the density, velocity and ram pressures would vary for a hot Jupiter orbiting any of these solar-like stars at a distance of 0.1 au. We see that the sun at minimum provides the lowest ram pressures of the sample (< 10$^5$ dyn cm$^{-2}$) while HD 190771 and $\chi^1$ Ori display the highest ram pressures with a maximum > 80$\times$10$^{-5}$ dyn cm$^{-2}$. This is useful for any further studies on planetary environment within the winds of G-type stars, with the age and rotation of the host star indirectly playing a role in the final ram pressure impacting the planets \dualta{and therefore upon atmospheric evaporation}. We examined how the velocities of these stellar winds are distributed globally, by taking a histogram of velocities at a distance of 0.1 au, shown in \Cref{fig:velocities}. We showed that more magnetically active stars display less uniform density distributions and overall have a more complicated structure.

We developed a numerical tool for calculating thermal radio emission from stellar winds given a simulation grid, removing the need for analytical formulations that have been used in the past \citep{Panagia1975,Fichtinger2017,Vidotto2017a}. This tool solves the radiative transfer equation for our wind models, which allowed us to derive radio flux densities, intensities and spectra. We found emission around the $\mu$Jy level with the winds staying optically thick up to 2 GHz. \dualta{We compared our calculated flux densities with recent observations and found our predictions agree with the observational upper-limits of $\kappa^1$ Ceti and $\chi^1$ Ori (F17 \& \citealt{Gaidos2000}).} Previous radio detections have been interpreted as originating in the chromospheres of solar-like stars and not their winds (F17 \& \citealt{Villadsen2014}), which is supported by our simulations. \par 
The normalised radio flux density emitted from these stellar winds is found to relate to stellar rotation as $S_{\rm \nu,1GHz}\propto\Omega^{0.7}$. This indicates that desired observational targets are stars with fast rotation rates within a distance of 10 pc. We showed in \Cref{fig:spectra} that more active close by stars like $\chi^1$ Ori and $\kappa^1$ Ceti would be readily detectable with the next generation of radio telescopes such as SKA and ngVLA.

\section*{Acknowledgements}
The authors wish to acknowledge the DJEI/DES/SFI/HEA Irish Centre for High-End Computing (ICHEC) for the provision of computational facilities and support. This work used the BATS-R-US tools developed at the University of Michigan Center for Space Environment Modeling and made available through the NASA Community Coordinated Modeling Center. D\'{O}F wishes to acknowledge funding received from the Trinity College Postgraduate Award through the School of Physics. AAV acknowledges funding received from the Irish Research Council Laureate Awards 2017/2018. SJ acknowledges the support of the  German Science Foundation (DFG) Research Unit FOR2544 ``Blue Planets around Red Stars'', project JE 701/3-1 and DFG priority program SPP 1992 ``Exploring the Diversity of Extrasolar Planets'' (RE 1664/18). We thank Jackie Villadsen and Joe Llama for their useful discussion on topics of stellar radio emission. The authors would like to thank our referee, Dr Jorge Zuluaga, for his valuable input on this work.
\\

\emph{Software:} \href{http://csem.engin.umich.edu/tools/swmf/}{BATS-R-US} \citep{Powell1999}, radiowinds \citep{radiowinds}, scipy \citep{scipy}, matplotlib \citep{matplotlib}, \href{www.tecplot.com}{TecPlot} and seaborn \citep{seaborn}.




\bibliographystyle{mnras}
\bibliography{main}




\appendix

\section{Effects of density and its gradient on radio emission}
\label{app:density}
\begin{figure}
    \centering
    \includegraphics[width=\linewidth]{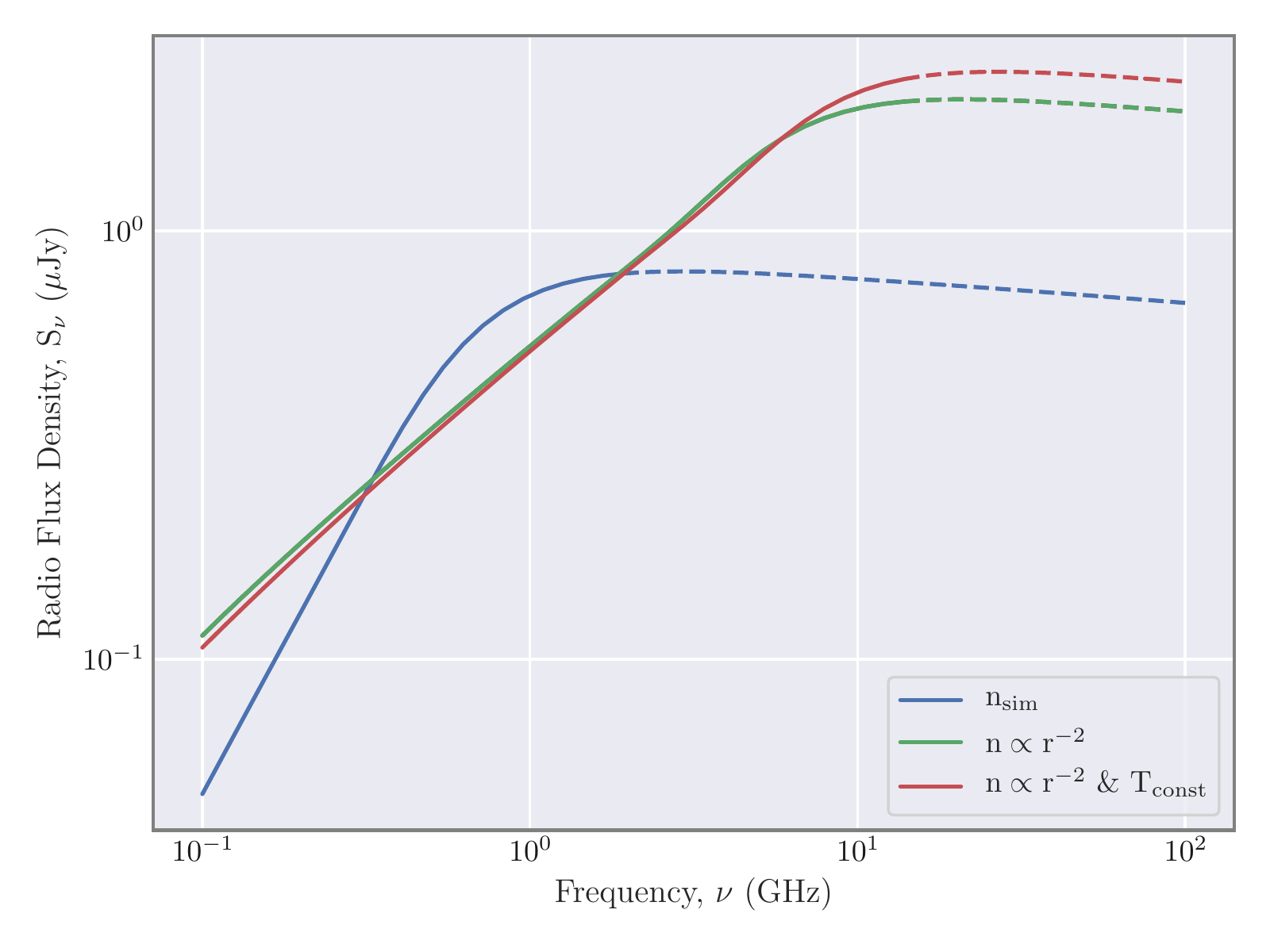}
    \caption{The blue line shows the same spectrum for $\kappa^1$ Ceti as shown in \Cref{fig:spectra}. The green line represents the same grid, with the same temperature gradient, but with a density that falls off with $R^{-2}$. The base density for this green line is also a factor of 3 smaller than the blue line. The red line represents the same scenario as the green line but with a constant temperature across the grid. Here we can see the huge impact density gradient has on flux density and spectrum shape.}
    \label{fig:spectrum_compare}
\end{figure}

Many previous analytical works have shown the strong dependence of thermal free-free radio emission on density gradients in the wind \citep{Panagia1975,Wright1975,Lim1996}. We show in \Cref{fig:spectrum_compare} how the flux density spectrum for $\kappa^1$ Ceti would change given a density gradient that follows $n \propto R^{-2}$ (green line), and in addition one that has a constant temperature (red line). Both of these models have a base density 3 times less than the original spectrum (blue line). We see that this slower density decay has a dramatic affect on the shape of the spectrum in the optically thick regime.  
The density gradient for our simulation varies across the grid, but in nearly all cases it is much steeper than $n \propto R^{-2}$. The steeper decay of density causes the emission to be lower across all frequencies. The temperature gradient has a minimal effect on spectrum shape compared to the density.

\begin{figure}
    \centering
    \includegraphics[width=\linewidth]{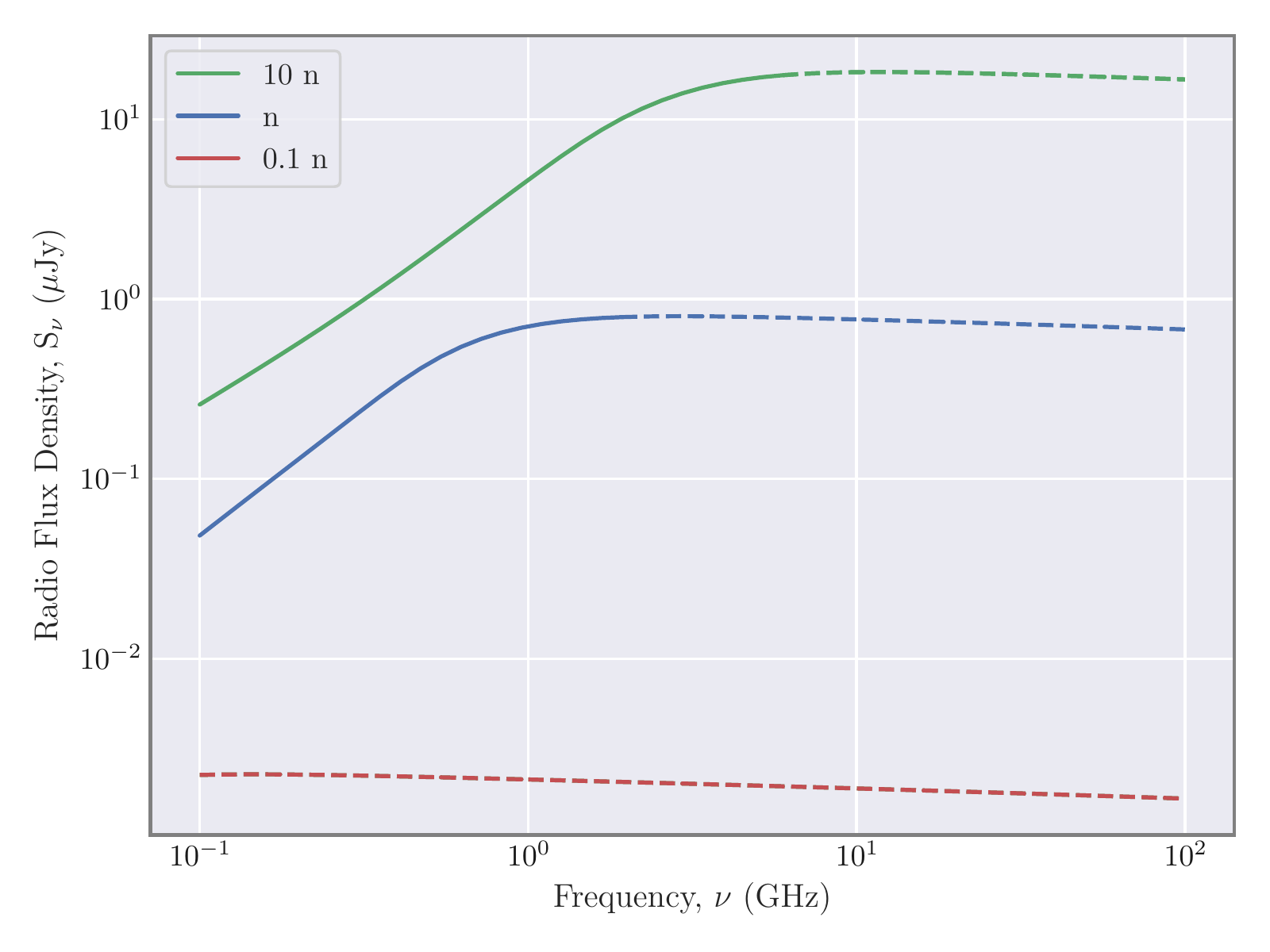}
    \caption{The blue line shows the same spectrum for $\kappa^1$ Ceti as shown in \Cref{fig:spectra}. The green line represents the same density structure with 10 times the original density, and the red line represents the original density divided by a factor of 10. The dashed portion of each line represents where the wind becomes optically thin. We see in the low density case that the entire wind is optically thin and emission is very low as there is an extremely tenuous wind. For the high density case we see much higher fluxes, and the wind is optically thick for most of the observing frequencies in our range. }
    \label{fig:change_n_spectrum}
\end{figure}

\Cref{fig:change_n_spectrum} shows how the density of the wind will affect the overall emission, changing where the wind becomes optically thick/thin, and the increase/decrease in the flux density. This is relevant to observations because, if two or more detections are made at different frequencies and follow the optically thin power law of $\propto \nu^{-0.1}$, then we can assume the wind is thin and therefore constrain the value for density in the wind. In the low density case the entire wind is optically thin and emission is very low as there is an extremely tenuous wind. For the high density case we see much higher fluxes, and the wind is optically thick for most of the observing frequencies in our range.

\bsp	
\label{lastpage}
\end{document}